\documentclass{aa} 
\usepackage{graphicx}
\usepackage{txfonts}
\usepackage{lipsum}
\usepackage{multirow}                           
\usepackage{lscape}            
                               
\usepackage{placeins}           
\usepackage{ulem}
\usepackage[dvipsnames]{xcolor}

\usepackage{threeparttable}
\usepackage[colorlinks=true,linkcolor=blue,citecolor=blue,urlcolor=blue]{hyperref}

\begin{document}
\defcitealias{Anjitha2025}{A25}
\defcitealias{Nakoneczny2021}{N21}
\defcitealias{Tinker2010}{T10}
\defcitealias{Comparat2017}{C17}

\date{}
\title{
Angular clustering and bias of photometric quasars\\in the Kilo-Degree Survey Data Release 4}
\titlerunning{Quasar photo-$z$s and clustering}

\author{Anjitha John William \and Maciej Bilicki \and Wojciech A. Hellwing \and Szymon J. Nakoneczny \and Priyanka Jalan  }

\institute{Center for Theoretical Physics, Polish Academy of Sciences, al. Lotnik\'{o}w 32/46, 02-668 Warsaw, Poland \thanks{\email{(anjithajm,bilicki,hellwing)@cft.edu.pl}}}
\authorrunning{John William et al.}

\abstract{
We investigate the angular clustering and effective bias of photometrically selected quasars in the Kilo-Degree Survey Data Release 4 (KiDS DR4). We update the previous photometric redshifts (photo-$z$s) of the KiDS quasars using Hybrid-z, a deep learning framework combining four-band KiDS images and nine-band KiDS+VIKING magnitudes. Hybrid-z is trained on the latest Dark Energy Spectroscopic Instrument (DESI) DR1 and Sloan Digital Sky Survey (SDSS) DR17 quasars matching with KiDS, and achieves average bias $\langle \delta z \rangle < 0.01$ and scatter $\sim 0.04(1 + z)$ on a test sample. 
The updated catalog of $\sim 157k$ quasars over $777~\mathrm{deg}^2$ is divided into four tomographic bins spanning $0.1 \leq z_{\mathrm{phot}} \leq 2.7$. In each bin, we measure the angular two-point correlation function and compare it with theoretical predictions for dark matter clustering. We estimate the best-fit scale-independent quasar bias, which increases from $b \approx 1.6$ at $z \approx 0.6$ to $b \approx 4.0$ at $z \approx 2.2$, and is well matched by a quadratic relation in redshift. Our clustering analysis indicates that KiDS quasars reside in dark matter halos of mass 
$\log_{10}(M_{\mathrm{eff}}/h^{-1}M_\odot)$ in the range $\sim 12.7$-$12.9$ and effective peak heights $\nu_{\mathrm{eff}}$ rising from $\sim 1.5$ to $2.9$ over our redshift span. We study two systematics that could affect the bias derivation: stellar contamination and the redshift distribution assumed in the theoretical modeling. The former has a negligible effect, whereas the latter significantly impacts the derived $b(z)$, emphasizing the importance of redshift calibration. Our work is the first cosmological application of quasars selected from KiDS and paves the way for future extensions in the final KiDS DR5, the Legacy Survey of Space and Time, or the 4-metre Multi-Object Spectroscopic Telescope.
}
\keywords{quasars: photometric redshifts - angular clustering - techniques: machine learning - catalogs: photometric \& spectroscopic }
\maketitle

\section{Introduction}

Quasars, also known as quasi-stellar objects \citep[QSO,][]{SCHMIDT_1963}, are the brightest among the 
active galactic nuclei \citep[AGN,][]{Rees1984}. 
Their high luminosity ($\sim10^{10} - 10^{14} L_\odot$) is due to the 
matter accretion on a supermassive black hole (SMBH) situated at the center of their host galaxy \citep{Salpeter1964, Zel'dovich1964}.  
This allows us to observe quasars 
up to very high redshifts ($z \gtrsim 7$),  which makes them valuable astrophysical and cosmological probes of both late and early-time Universe \citep{Fan2023}.
{ Quasar emission originates from regions well within the gravitational influence of the supermassive black hole (SMBH), providing information on the physics of black hole accretion and AGN activity \citep{Marziani2006, Spilker2025}. At galactic scales (\(\sim\)1–10 kpc), quasars help in understanding galaxy formation, growth, and quenching processes \citep{Prochaska2009}. On cosmological scales ($\gtrsim 1$ Mpc), they trace the large-scale structure (LSS) of the Universe \citep{Song2016}.
For instance, measuring quasar clustering allows us to constrain the redshift evolution of their bias with respect to the underlying dark matter (DM), reveal their host halo masses, and provide access to cosmological information such as detecting baryon acoustic oscillations \citep[e.g.][]{Myers2007a, Ata2018, Adame2025}. 
Their spatial distribution can be cross-correlated with galaxy and cosmic microwave background weak gravitational lensing maps to probe the growth of structure and constrain cosmological parameters \citep[e.g.][]{Bartelmann2002, DiPompeo2016, Alonso2023, Luo2024, Piccirilli2024}. Additionally, the large-scale clustering of quasars is sensitive to primordial non-Gaussianity through scale-dependent bias effects \citep{Slosar2008}.}

Despite quasars being among the brightest extragalactic objects, selecting them in surveys remains challenging. Their compact emission region outshines the host galaxy, and at high redshifts, their small angular size, combined with the limited resolution of telescopes, causes them to appear point-like, making them morphologically indistinguishable from stars in imaging data. As a result, quasar selection must rely on other properties, such as colors, variability, or spectroscopy, to separate them from the vastly more numerous stellar population. 
However, the spectroscopic quasar identification method \citep{Richards2002, Lyke2020,Storey2024} 
is time-consuming and observationally expensive, restricting surveys to relatively bright sources and limiting sky coverage and depth of quasar samples. {To overcome} these limitations, we can apply photometric selection techniques to produce larger and more complete QSO catalogs \citep[e.g.][]{Carrasco2015, Nakoneczny2019,Chaussidon2023, Feng2025}.
Photometrically classified quasars {offer higher completeness but at the cost of increased contamination } compared to their spectroscopic counterparts.
However, they provide substantial cosmological information due to their large numbers and wide sky coverage.

{
Photometric target selection technique for quasar identification is applied within the Kilo-Degree Survey \citep[KiDS,][]{Kuijken2019} by \citealt[][hereafter \citetalias{Nakoneczny2021}]{Nakoneczny2021}. KiDS covers about 1000 deg$^2$ of the southern sky in its fourth data release.
This paper presents the first study of angular clustering and {effective} bias for KiDS DR4 quasars.
Our results demonstrate that the quasar bias evolution is consistent with a quadratic dependence on redshift, in agreement with prior findings from other surveys \citep{Myers2007a, Shen2009, Laurent2017,Eltvedt2024}.  
We investigate potential systematic effects, finding that uncertainties in the redshift distribution significantly impact the quasar bias estimation, highlighting the necessity of precise redshift calibration. In contrast, stellar contamination within the sample has a negligible effect on the clustering measurements and bias inference. Together, these findings establish a framework for quasar clustering studies with KiDS and contribute insight into quasar bias evolution across cosmic time.

A prerequisite to study the evolution of quasar properties -- such as their bias -- is the availability of redshift measurements. Similar to non-active galaxies, the redshifts of quasars can be determined through both spectroscopic and photometric methods. 
Measuring the accurate and precise spectroscopic redshifts (hereafter spec-$z$s) is, however, both expensive and time-consuming. The less precise photometric redshift (photo-$z$) estimation is based on broadband photometry \citep{Baum1957, Koo1985,Salvato2019, Newman2022}. 
This alternative approach can be relatively quick, as compared to spectroscopy, and may give a redshift estimate for all sources within an imaging survey. 
In this work, we use the empirical approach of machine learning (ML) to derive photo-$z$s for KiDS DR4 quasars. 
Such algorithms, typically belonging to the supervised ML category, find the 
relation between multi-band photometric quantities and the redshift by being trained on datasets containing both photometry and corresponding spec-$z$s.
Examples of their applications to quasars include 
\cite{Brescia2013,DIsanto2018, Nakoneczny2021, Nakazono2024}.

As compared to other galaxies, quasars exhibit prominent broad emission lines, such as \(\mathrm{Ly \alpha}\), C\,IV, Mg\,II \citep{Peterson1997}, and they 
move across different photometric bands at different redshifts.
Accurate photometric redshift estimation for quasars therefore benefits from broad wavelength coverage, extending from the blue to the infrared (IR), to help break degeneracies caused by different emission lines entering similar filter combinations at different redshifts, as well as ambiguities from their smooth, power-law continua \citep{Salvato2019}. Still, these effects often produce characteristic peaks and dips in the quasar photometric redshift distribution. 
Furthermore, QSO spectra are dominated by non-thermal emission, giving them colors distinct from most other extragalactic sources and necessitating training sets specifically tailored for quasar photo-$z$ measurements.

Within the broader ML framework, {deep learning (DL), often using images among the inputs, has been finding increasingly numerous uses for photo-$z$ estimation, in particular for quasars \citep[e.g.][]{PasquetItam2017, Yao2023,Roster2024}.
The term ``deep'' indicates artificial neural networks (ANN) with multiple layers (often tens or hundreds of them concatenated) and numerous neurons in each layer. 
In this work, we integrated two types of neural networks, dense \citep {mcculloch1943, Rosenblatt1958} and convolutional \citep[CNN,][]{Lecun1995}. 
Our model is called Hybrid-z\footnote{\url{https://hybrid-z.readthedocs.io/en/latest/}} since it uses both magnitudes and multi-channel images, {and was originally developed for the KiDS bright galaxy sample in} \citealt[][hereafter \citetalias{Anjitha2025}]{Anjitha2025}. {In its previous application,} Hybrid-z improved the earlier ANN-based photo-$z$s for the KiDS DR4 bright sample \citep{Bilicki2021}{, reducing the scatter of photo-$z$ residuals by 20\%}. In this work, we apply a similar DL model to update the photo-$z$s of the photometrically classified KiDS DR4 QSO sample, previously derived with ANNs employing quasar magnitudes and colors by
\citetalias{Nakoneczny2021}.

In addition to developing a new ML model for photo-$z$s, a key improvement of this work over \citetalias{Nakoneczny2021} is the use of the recent Dark Energy Spectroscopic Instrument Data Release 1 \citep[DESI DR1,][]{desidr12025}
and Sloan Digital Sky Survey Data Release 17 \citep[SDSS DR17,][]{Abdurrof2022} to build the spectroscopic training set for
our photo-$z$ model. As compared to the SDSS DR14 \citep{Abolfathi2018} spec-$z$ data used previously by \citetalias{Nakoneczny2021}, adding DESI DR1 considerably improves the quality of our training set and the resulting photo-$z$ estimation. DESI not only includes many more quasars than SDSS in the KiDS fields; it also extends the QSO spectroscopic coverage across the color-redshift space.

We use the KiDS DR4 quasar sample with the updated photo-$z$s to study the connection between QSO distribution and the underlying DM field via angular clustering, which is a common observational statistic to study the properties of the LSS projected on the sky \citep{Peebles1973,Peebles1980}. Following previous studies \citep[e.g.][]{Myers2007a}, we consider quasar auto-correlations in photo-$z$ bins.
We use these measurements to constrain the evolution with redshift of the effective quasar bias and of their effective host halo mass.

In the standard cosmological framework, the evolution of density perturbations yields predictions for the power spectrum of the total matter density contrast field, including contributions from both dark and baryonic matter. 
Observationally, we do not directly probe the clustering of the total matter density field. Instead, we measure the spatial distribution of luminous tracers
- {in our case, quasars. On sufficiently large scales (megaparsecs), the DM and quasar overdensities ($\delta_\mathrm{q}$) are related via the  bias parameter, such that $\delta_\mathrm{q} = b \delta_\mathrm{m}$. This bias is expected to evolve with redshift 
\citep{ Shen2009,Laurent2017} and its evolution can be traced from angular clustering measured 'tomographically', i.e., in redshift bins. If additional information on quasar redshift distribution per bin is available -- i.e., via photo-$z$s -- the angular two-point correlation function (2PCF) can be related to the theoretical three-dimensional 2PCF or its Fourier counterpart, power spectrum. The factor linking the observed quasar correlations and those theoretically expected for DM is the bias $b$. {In summary, incorporating the redshift distribution enables the projection of theoretical three-dimensional dark matter clustering into angular space, allowing the quasar bias to be inferred from the observed 2PCF.}

This paper is structured as follows: In Sec.\ref{Data}, we describe the photometric and spectroscopic datasets used in our analysis. Sec.\ref{sec:photo-z method} presents our Hybrid-z model and the photo-$z$ estimation for the quasar sample. The clustering analysis of this sample is detailed in Sec.\ref{sec:clustering}. Finally, we summarize our findings and discuss future prospects in Sec.\ref{sec:conclusion}.

\section{Data}
\label{Data}

In this section, we describe the data we use to select quasars and estimate their photo-$z$s.
The quasar sample is derived from {KiDS Data Release 4 \citep[DR4,][]{Kuijken2019}}, and the spectroscopic data for the training sample are obtained from {DESI DR1 and SDSS DR17}. 

KiDS is a multi-band imaging survey designed for weak lensing studies and it was carried out by the European Southern Observatory (ESO) at the VLT Survey Telescope \citep[VST,][]{Capaccioli2011}. The images were taken in four optical broad bands (\(\mathrm{ugri}\)) {and supplemented with five near-infrared bands ($\mathrm{ZYJHK_s}$) from the VISTA Kilo-degree INfrared Galaxy survey \citep[VIKING,][]{Edge2013} overlapping with KiDS on the sky. Here we employ KiDS DR4, covering about 1000 deg$^2$ before masking. The nine-band images were processed by the KiDS team, using in particular the {Gaussian aperture and point spread function method} \citep[GAaP,][]{Kuijken2008} to obtain uniformly measured $u$ to $K_s$ magnitudes, which we employ throughout. In addition to the magnitudes, similarly as in \citetalias{Anjitha2025}, for photo-$z$ derivations we also use four-band $\mathrm{ugri}$ images, leaving the extension to VIKING images for future work.}   
{The optical images in KiDS DR4 are organized into 4$\times$1006 survey tiles\footnote{\url{https://kids.strw.leidenuniv.nl/DR4}} with a uniform pixel scale of 0.2 arcsec \citep{deJong2015}. 
For the Hybrid-z model described below, we 
made cutouts centered at quasar positions with a size of $5'' \times 5''$  ($25 \times 25$  pixels), as most of our quasars are smaller than this.

Our quasar sample is selected from KiDS DR4 following the previous work by \citetalias{Nakoneczny2021}\footnote{\url{https://kids.strw.leidenuniv.nl/DR4/quasarcatalog.php}}. 
That selection was obtained by applying an ANN-based classification model using 9-band detections, trained and tested on SDSS DR14 spectroscopy.  
In the KiDS DR4 QSO sample, the objects are further categorized into ``safe'' and ``extrapolation'' regimes based on the feature space coverage with respect to the training set. {In this work, for the clustering analysis presented in Sec.~\ref{sec:clustering}, we only use the safe subset, limited to $r<22$ mag,} whose feature space lies well within the domain of the \citetalias{Nakoneczny2021} training data.  For this subset, the \citetalias{Nakoneczny2021} classification model achieves test-data purity of 97\% and completeness of 94\%. As we do not update the quasar selection over that previous study, limiting our analysis to only the safe QSO set mitigates potential contamination from stars and non-active galaxies, which could affect photo-$z$ quality and subsequent clustering analyses. There are $\sim157k$ KiDS DR4 objects in this safe QSO category. Unless otherwise specified, the term KiDS DR4 quasar sample hereafter refers to this subset of safe quasars. Note, however, that the general training set does not have the safe selection applied (see below). We leave for future study a possible analysis of quasars derived from KiDS DR5 \citep{Wright2024,Feng2025}.
}

While we adopt the same KiDS DR4 QSO sample as selected by \citetalias{Nakoneczny2021}, we update and improve the photo-$z$s derived there. This is achieved by both changing the methodology (ML photo-$z$ model) as well as by adding considerably more training data with spectroscopic labels. The former is provided thanks to using the Hybrid-z approach, employing both 9-band magnitudes and 4-band optical images, previously developed and validated on the KiDS-bright galaxy sample \citep{Bilicki2021} by \citetalias{Anjitha2025}. We discuss the particular Hybrid-z implementation and its performance for KiDS DR4 quasars in the next Sec. \ref{sec:photo-z method}.

{As for the new training data, we replace the SDSS DR14 QSOs previously used by \citetalias{Nakoneczny2021} by a joint sample of SDSS DR17 and DESI DR1 quasars. These two datasets cover the northern (equatorial) KiDS area and provide {spectroscopically confirmed quasars with} spectroscopic redshifts {(that may be considered exact)} to train and validate our photo-$z$ model. The joint SDSS+DESI sample, with only unique objects kept\footnote{Note that majority of SDSS quasars are also present in the DESI catalog. In the cross-match with KiDS DR4, only about 1900 out of 17,000 SDSS QSOs are not in DESI.}, is dominated by DESI and has about 115k counterparts in the full KiDS DR4 catalog before we apply any further cuts. Of these, about 104k have all nine $\mathrm{ugriZYJHK_s}$ bands measured in KiDS DR4 -- a selection we apply as our neural network model requires a numerical value to be provided for each of the magnitudes. This general cross-match of DESI+SDSS with KiDS DR4 
will be our training and test set for the Hybrid-z model. Note that this training sample goes deeper in magnitude and redshift than the final safe-QSO sample we use for the clustering analysis, which is a conscious choice to provide photo-$z$ estimates also for quasars fainter than eventually employed.
The median $r$-band magnitude of the quasar sample with spectroscopic redshift labels is $r\sim21.4$ mag, while the depth expressed as the 99th percentile is $r\sim23.7$. The median spectroscopic redshift of this training set is $z=1.66$ and 99\% of the quasars are contained within the {redshift range} $0.22\leq z \leq 3.32$.}

\section{Photometric redshift derivation}
\label{sec:photo-z method}

The architecture and configuration of our photo-$z$ model, Hybrid-z, are detailed in \citetalias{Anjitha2025}. In brief, 
Hybrid-z comprises two parallel branches: a CNN branch that processes \(\mathrm{ugri}\) KiDS galaxy cutouts through initial convolutional layers and four Inception modules to extract multi-scale features, and an ordinary (fully-connected) neural network (ONN) branch that processes standardized nine-band KiDS+VIKING magnitudes through a sequence of dense layers. The CNN and ONN outputs are concatenated and passed through fully connected layers, with the final layer providing a photo-$z$ estimate.
The primary modification between the current model configuration and that detailed in \citetalias{Anjitha2025} is the exclusion of the sigmoid activation function in the output layer.
This adjustment was made since the redshift distribution of quasars within our dataset extends well above $z>1$, while previously we were limiting
the predictions to the $0<z_\mathrm{phot}<1$ range, appropriate for the KiDS-bright sample. Therefore, here a linear activation function is applied in the final layer. The rest of the Hybrid-z architecture remains identical to that of \citetalias{Anjitha2025}.
{The hyperparameters differ from those in \citetalias{Anjitha2025} and were selected through empirical testing across various configurations and evaluating the corresponding model's performance on the quasar dataset used in this study.}
We employed a batch size of 32 and an initial learning rate of 0.001.

As far as data preprocessing is concerned, we adapt the same techniques as in \citetalias{Anjitha2025}; normalization to the image pixel values and standardization of the 9-band magnitude features. 
Furthermore, we employed data augmentation, including $90^\circ$
rotation, flipping, and translations (height and width shifts), to  expand our training set \citep{yang2022}.
We used $\zeta=\ln(1 + z_\mathrm{spec})$  as labels \citep{Baldry2018} to achieve a more balanced distribution in redshift. This transformation of true values helps to maintain stable gradients in the  descent-based optimization of the model by ensuring numerical stability. 
Although the model is trained in this log-transformed space, all evaluations and error metrics are reported after inverting the transformation.

The dataset is divided into training, validation, and testing disjoint sets using the \textsc{scikit-learn} library function \verb|train_test_split|, with approximately two-thirds of the data assigned to training and the remainder evenly split between validation and testing subsets. In our case, this corresponds to $\sim$104k quasars in total, with $\sim 73k$ allocated for training and $\sim 18k$ each for validation and testing.  

The performance of the Hybrid-z model is primarily evaluated using the Huber loss function \citep{huber}. The loss curves for both training and validation sets show a steady decline over successive epochs, indicating efficient learning and convergence. The close agreement between the two curves suggests that the model generalizes well to unseen data without signs of overfitting. We used early stopping criteria \citep{Prechelt1996} to determine the optimal number of epochs. Training ceased automatically after 10 consecutive epochs without improvement, with any reduction in validation loss considered as an indication of improvement. The model was trained over 70 epochs in our model setup. 
Finally, the estimated QSO photo-$z$s for the test sample are evaluated by standard statistical metrics. 
To quantify the statistical accuracy, we calculated the bias, \(\delta z= z_\mathrm{phot} - z_\mathrm{spec}\), and normalized bias, \(\Delta z= \frac {\delta z}{1+z_\mathrm{spec}}\). Standard deviation of normalized bias $\sigma_{\Delta z}$ and scaled median absolute deviation (SMAD) of $\Delta z$ are the measures of scatter, i.e. statistical precision. Here, \(\mathrm{SMAD}(\Delta z) = 1.4826 \times \text{median}(\lvert \Delta z - \text{median}(\Delta z) \rvert)\).

\subsection{Results for test data }

\label{sec:test data}

\begin{figure}[!htbp] 
    \centering
        \includegraphics[width=0.5\textwidth]{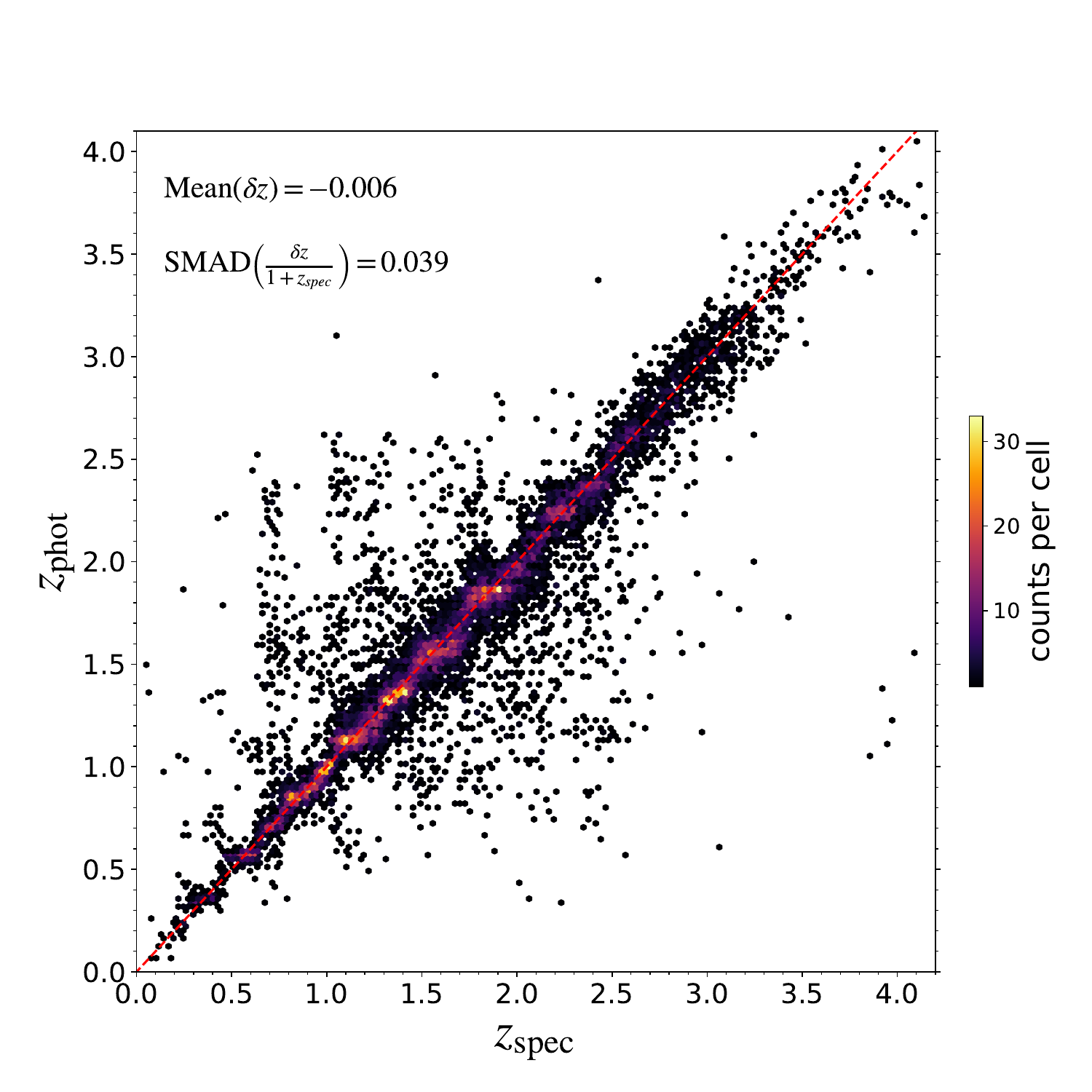}
        \caption{
Density plot comparing spectroscopic and photometric redshifts of KiDS DR4 quasars derived with our Hybrid-z method. 
The color bar indicates counts per hex-bin {for a test sample not seen in the training}. 
The red dashed line corresponds to the identity relation $z_{\mathrm{spec}} = z_{\mathrm{phot}}$.
}
        \label{fig:test_data}
\end{figure}

\begin{table*}
\begin{threeparttable}

\caption{Photometric redshift statistics of the Hybrid-z model for quasars in a 
test sample {derived from a cross-match of KiDS DR4 with DESI+SDSS spectroscopy. Row 1 applies to the entire test sample (not seen by the model in training). Rows 2-3 are for a cross-match with the earlier \citetalias{Nakoneczny2021} derivations: original results in row 2, and our updates in row 3. Finally, rows 4-7 present a break-down into }  photometric redshift bins which are employed for tomographic angular clustering in this paper}. 
\label{table:test_data}

\small

\begin{tabular}{l c c l c c c c }

\hline\hline\\

Selection & Size \tnote{1} & {$\langle z_\mathrm{spec}\rangle$}  & \multicolumn{1}{c}{$\langle z_\mathrm{phot} \rangle$} & \multicolumn{1}{c}{ $\langle \delta z\rangle$} & $\langle \Delta z\rangle$ & \multicolumn{1}{c}{$\sigma_{\Delta z}$} & \multicolumn{1}{c}{\(\mathrm{SMAD}(\Delta z)\)} \\
\\
\hline\hline\\
    full test set\tnote{2}&18,274&1.662&1.656 &-0.006 &0.008 &0.133&0.039\\
    
    \multicolumn{5}{l}{test set cross-matched with the \citetalias{Nakoneczny2021} sample:\tnote{3}}\\
        {\qquad previous \citetalias{Nakoneczny2021} derivations\tnote{4}} &10,226 &1.672&1.738&0.067&0.038&0.131&0.054\\
        {\qquad this work} &  10,226 & 1.672  &1.664 & -0.007 & 0.005 & 0.116 &0.033 \\
      
      \\
      \hline
      \\
      $0.1 \leq z_\mathrm{phot} \leq 0.8$  &718 &0.656 &0.596& -0.061& -0.019&0.104&0.029\\
      $0.8 < z_\mathrm{phot} \leq 1.2$ & 1996& 1.079&1.018&-0.061&-0.017&0.095&0.030\\
       $1.2 < z_\mathrm{phot} \leq 1.8$ &3459&1.510&1.489&-0.021&0.002&0.110&0.038\\
       $1.8 < z_\mathrm{phot} \leq 2.7$        &3290&2.110&2.158&0.047&0.027&0.139&0.031\\

\\
\hline 
\hline

\end{tabular}
\begin{tablenotes}
\item[1] Number of quasars in the subsample.
\item[2] Entire spectroscopic test set without cross-matching to \citetalias{Nakoneczny2021} selection nor any cuts.
\item[3] Spectroscopic test set cross-matched with the \citetalias{Nakoneczny2021} safe quasar sample ($r<22$ mag).
\item[4] \citetalias{Nakoneczny2021} derived photo-$z$ statistics in the spectroscopic test set cross-matched with the \citetalias{Nakoneczny2021} safe quasar sample ($r<22$ mag).
\end{tablenotes}

\end{threeparttable}
\end{table*}

In this section, we analyze the KiDS DR4 QSO photo-$z$s derived using the Hybrid-z model for the test sample, with true redshift known, but not seen by the model in training nor validation.} 
Starting with the statistics averaged over the entire test set {of about 18,200 quasars drawn from a general cross-match of KiDS with DESI+SDSS spectroscopy,} the mean bias is below 0.01 in absolute terms (see first row of Table~\ref{table:test_data}), while the scatter of $\Delta z \equiv (z_\mathrm{phot}-z_\mathrm{spec})/(1+z_\mathrm{spec})$ is
$\sigma_{\Delta z} = 0.133$ {(standard deviation)} or $\mathrm{SMAD}({\Delta z})=0.039$ {(scaled median absolute deviation).} 
Therefore, our photo-$z$s are on average unbiased i.e., the overall scatter is considerably larger than the mean bias. On the other hand, the fact that SMAD is much smaller than SD indicates their non-Gaussian nature with extended tails (outliers), which is also visible in the direct $z_\mathrm{spec}$ -- $z_\mathrm{phot}$ comparison, presented in Fig.~\ref{fig:test_data}, as estimates far apart from the diagonal.
This is typical behavior for quasar photo-$z$s \citep[e.g.][]{Curran2022, Yao2023}. 

In order to directly compare the Hybrid-z performance for KiDS DR4 QSOs with the previous derivations by \citetalias{Nakoneczny2021}, we cross-matched the above-mentioned test sample with the safe set from that earlier work. This reduces the number of common objects to about 10k mostly because of the $r<22$ mag limit in that sample. The results are presented in rows 2-3 of Table~\ref{table:test_data} and we notice that 
the Hybrid-z model exhibits a notable improvement {over \citetalias{Nakoneczny2021} in all the statistics.}
As we use the \citetalias{Nakoneczny2021} safe quasar sample for clustering measurements below, the statistics provided in row 3 of Table~\ref{table:test_data} us general quantification of the photo-$z$ performance for our dataset.

}

{We illustrate a direct comparison of our photo-$z$ predictions for the test sample with the true ones in Fig. \ref{fig:test_data}. While the data generally follow the diagonal (identity line $z_\mathrm{spec}=z_\mathrm{phot}$), as expected for photo-$z$s unbiased on average, characteristic `step-like' behavior is observed in detail. This specific redshift focusing, where a range of spec-$z$s have similar photo-$z$s predicted, or vice versa, is  
commonly observed in quasar photo-$z$ derivations \citep[e.g.][]{Nakoneczny2021,KunsagiMate2022, Yao2023,Moss2025}.
{We interpret these steps as artifacts driven by emission line transitions across filters.} 
For instance, At lower redshifts (e.g., $z \sim 0.4$), the H$\alpha$ emission line (restframe $\lambda = 6562.8\AA$) lies within the $Z$-band {(central $\lambda =  8800\AA$)}, contributing significantly to redshift estimation. As redshift increases, H$\alpha$ shifts out of that filter coverage, and although H$\beta$ remains visible, this transitional zone introduces increased uncertainty. Around $z \sim 1.1$, H$\beta$ {($4861\AA$)} also exits the optical range, and the model increasingly relies on redshifted Mg\textsc{ii} {($2800\AA$) and C\textsc{iii]} ($1908.73\AA$) emission lines}, which can lead to degeneracies due to similar observed colors across different redshifts. These transitions result in redshift confusion, where the model tends to assign many objects the same redshift, producing the observed step-like artifacts in the distribution. Each ``step'' reflects a regime where the model’s redshift estimation is dominated by a different combination of emission lines.

The KiDS quasar clustering analysis presented below in Sec.~\ref{sec:clustering} is done tomographically in four photo-$z$ bins, it is therefore relevant to analyze the redshift performance for each of these bins. This is presented in rows 4-7 of Table.\ref{table:test_data}{, where the numbers refer to a cross-match of our test set with the \citetalias{Nakoneczny2021} safe quasar sample, binned in Hybrid-z redshift estimates. This time we observe more considerable biases in our photo-$z$s, with mean $\delta z = z_\mathrm{phot} - z_\mathrm{spec}$ surpassing 0.01 in absolute terms, with some improvement if the $1+z$ scaling is applied (i.e. for $\Delta z$). On the other hand, the scatter is better behaved, staying at a similar level for the particular bins as for the full test sample, both in terms of the standard deviation and SMAD of the residuals. This indicates that, once binned in photo-$z$, our redshift estimates retain their statistical precision, but lose on accuracy.}

\subsection{Updated KiDS DR4 quasar photometric redshifts}
\label{sec:kids_dr4_qso}

Having checked the Hybrid-z performance for quasars with known redshifts from spectroscopic data, we trained the model on the KiDS DR4 $\times$ DESI+SDSS cross-match and applied it to the entire safe quasar sample as selected by \citetalias{Nakoneczny2021}. As already clarified, our feature space are 4-band $\mathrm{ugri}$ images and 9-band KiDS+VIKING magnitudes, with a requirement of having detections in each. This dataset, flux-limited to $r<22$ mag, includes 157,419 quasar candidates on $\sim777$ deg$^2${(effective sky coverage of KiDS DR4 after masking)}. In Fig.~\ref{fig:redshift_comparison} we compare the redshift distribution of the spectroscopic training quasars present also in our photometric set (filled histogram) with two photometric redshift distributions: previous from \citetalias{Nakoneczny2021} (green dashed) and the one derived here with Hybrid-z (blue solid). Each of the histograms was normalized to unity under the curve. The comparison indicates that the original relatively smooth $dN/dz_\mathrm{spec}$ is mapped to $dN/dz_\mathrm{phot}$ with visible peaks and dips, both for \citetalias{Nakoneczny2021} and Hybrid-z derivations. This is another manifestation of the effects already discussed previously in the context of Fig.~\ref{fig:test_data} that quasar photo-$z$s tend to be focusing around particular redshift values due to redshifted line transitions. This is a caveat one should bear in mind when analyzing the overall photo-$z$ performance for quasars. Indeed, as shown above in Table \ref{table:test_data}, once broken down into particular redshift bins, the statistics may deteriorate. On the other hand, for the subsequent clustering analysis we chose sufficiently broad photo-$z$ ranges to include neighboring `peaks' and `dips' in the redshift distribution and hence partly mitigate the related issues.

\begin{figure}[!htbp] 
    \centering
    \includegraphics[width=0.5\textwidth]{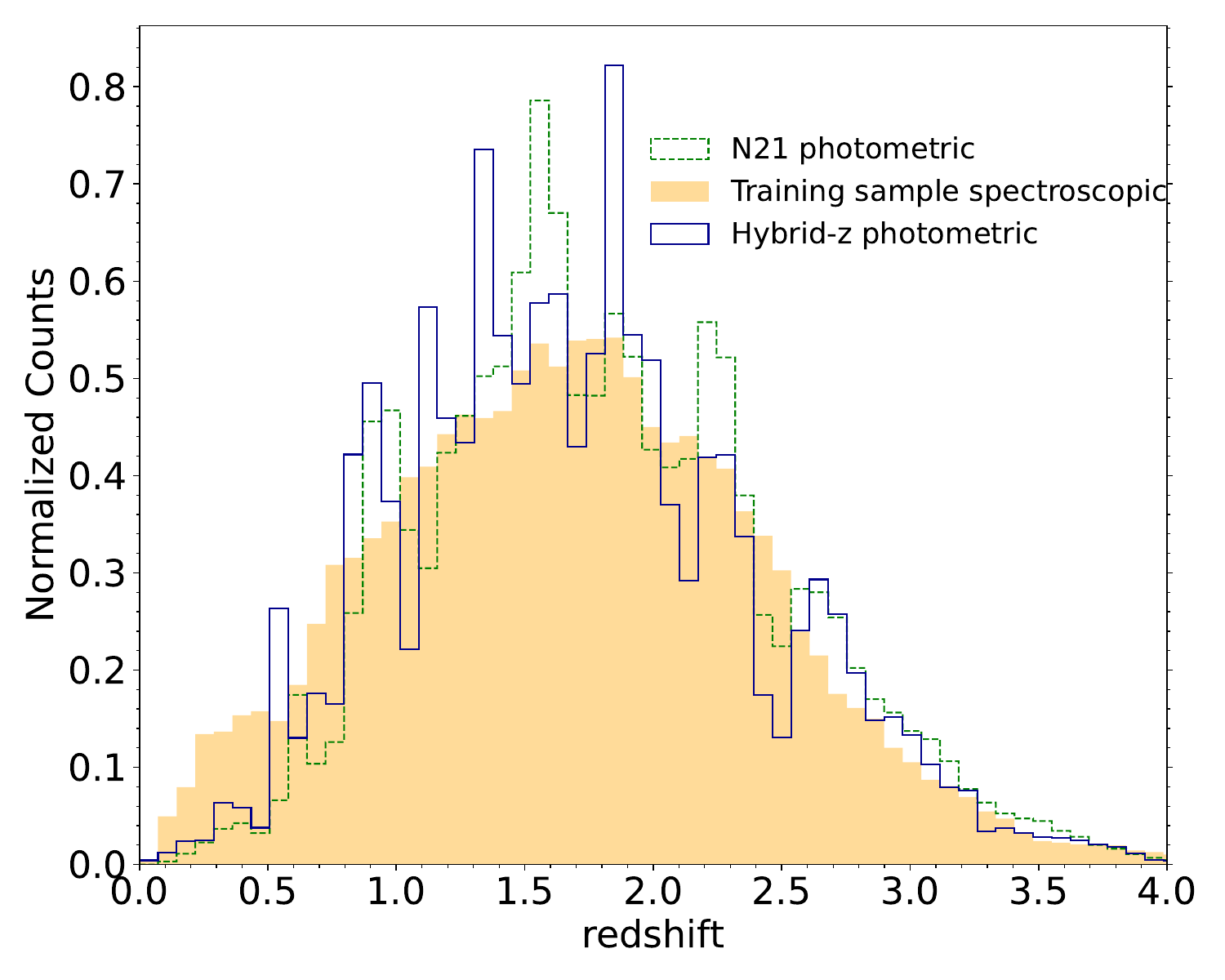}
\caption{{Comparison of redshift distributions for the KiDS DR4 safe quasar sample. The orange-filled histogram shows the spectroscopic training sample used in our Hybrid-z {model}. The green dotted line represents the photo-$z$s of KiDS DR4 safe quasars from \citetalias{Nakoneczny2021}, while the blue line corresponds to the photo-$z$s derived in this work using the Hybrid-z framework.} 
}
    \label{fig:redshift_comparison}
\end{figure}

\section{Angular clustering and bias evolution of KiDS quasars}
\label{sec:clustering}

{In this section, we study the clustering of KiDS quasars along with their effective bias, peak height and host halo mass evolution. Our observable is the angular 2PCF measured in photo-$z$ bins and compared to theoretical predictions for the underlying DM field. We also examine two possible systematics -- stellar contamination and redshift distribution modeling.

\subsection{Measurements: angular correlation function}
\label{sec:observed}

We use the angular two-point correlation function (2PCF), $\omega(\theta)$, as our observable. The 2PCF quantifies the excess probability of finding a pair of quasars separated by a given distance, here expressed as an angle $\theta$ on the sky \citep{Peebles1973}, as we are dealing with photometric data without exact redshifts. To compute the 2PCF, we use the standard \cite{Landy1993} estimator:
\begin{equation}
    \omega_{obs}(\theta) = \frac{DD (\theta) - 2 DR(\theta)+ RR(\theta)}{RR(\theta)}\;.
\end{equation}
Here, \(DD(\theta)\), \(DR(\theta)\), and \(RR(\theta)\) represent the 
pair counts normalized by number density within an angular bin centered at $\theta$, for the data-data, data-random, and random-random pairs, respectively. The random catalog covers the same footprint as the real data and includes about 100x more points than the quasar sample, for each of the redshift bins used. It is generated by first uniformly distributing points within the range of the minimum and maximum values of right ascension and declination of the data and then applying the KiDS DR4  binary HEALPix mask with \( N_{\text{side}} = 4096 \) 
to restrict the random dataset to the survey coverage. In this work we do not employ any corrections to the randoms that would compensate for survey systematics. While appropriate methodology has been developed for KiDS \citep{Johnston2021,Yan2025}, our quasar sample is too sparse for such `organized randoms' approach to be applicable. 
{Position-dependent variations of survey depth could introduce systematic effects in clustering measurements, which usually lead to the amplitudes being overestimated \citep[e.g.][]{Johnston2021,Vakili2023,Yan2025}. However, our sample is limited to $r < 22$~mag, which is well below the KiDS limiting magnitude of $r \sim 25$~mag. {In addition, we work with point sources, which should be less sensitive to effects such as varying PSF.} Therefore, we do not expect the effective depth to be a dominant source of uncertainty for our analysis.}

Our measurements are performed in nine logarithmically-spaced angular bins in the range $0.05^\circ < \theta < 1^\circ$. The lower value is driven by shot noise effects in our relatively sparse dataset, while the upper one  corresponds to very large physical scales once deprojected from angles to distances, and additionally is chosen to mitigate the main KiDS systematic, which is number density variations between tiles of 1 degree at a side. We have checked that indeed the measured $\omega(\theta)$ displays `jumps' when passing to larger scales, hence we discard them in the analysis. Restricting measurements to scales below $1^\circ$, while necessary, will of course not remove the systematics arising from tile-to-tile variations, as correlations between quasars near tile edges will still be present. However, similarly as with the general effect of variable depth, we assess this to be of less importance here as our sample covers the bright end of KiDS.
As for small scales, while we do measure the correlations down to arcminutes, the first angular bins are not used to derive the best-fit quasar bias when comparing with theoretical predictions. This is because the corresponding physical scales fall within the non-linear regime where our assumed simple relation between matter and quasar clustering is no longer adequate.

In our analysis we employ four photo-$z$ bins with edges $z_\mathrm{phot}=\{ 0.1,0.8,1.2,1.8,2.7 \}$ (Table.\ref{table:test_data}). The minimum and maximum values were chosen to cover the bulk of the redshift distribution (Fig.~\ref{fig:redshift_comparison}), while the bin widths and divisions are selected to account for photo-$z$ scatter, quasar number density and the artifacts in the redshift distribution, as discussed in Sec.~\ref{sec:kids_dr4_qso}. The angular 2PCF is computed separately in each photo-$z$ bin using the $\tt{TreeCorr}$ package \citep{Jarvis2004}, and the measurements are shown as points in Fig.~\ref{fig:correlation}. The number of sources in each redshift bin is indicated in the legend of each panel in Fig.~\ref{fig:correlation}. The error bars attached to the measurements in Fig.~\ref{fig:correlation} are the square root of the diagonal elements of the covariance matrix (Sec.~\ref{sec:covariance}).
Generally, the larger error bars in the first redshift bin than those in the other three bins are primarily driven by the smaller number of sources and the increased relative impact of shot noise in that bin.

{In this work, we use only auto-correlations of quasars in the redshift bins. However, due to photo-$z$ errors, there is also  clustering signal in bin cross-correlation \citep{Blake2006, Agarwal2014, Ho2015}, notwithstanding the magnification lensing effect correlating the bins even for exact redshift tomography \citep{Breton2022}. What is more, photo-$z$ bin cross-correlations could be used as a further diagnostic for systematics, for instance, in case of redshift outliers \citep{Balaguera2018}. However, in this work we chose to focus on auto-correlations only for two main reasons. First, proper modeling of cross-correlations would require deeper knowledge of the underlying true redshift distributions for the bins, which we do not have. Second, such modeling would need to properly account for lensing magnification, which is usually quantified by assuming magnitude-limited sampling \citep{Maartens2021}. While our quasar selection does have a flux cut applied, it is far from magnitude-limited due to complexities of ML-based catalog construction. We are therefore not able to properly model the cross-correlation measurements to either use them as extra information or to evaluate residual systematics.}

\subsection{Theoretical modeling}
\label{sec:theoretical_modeling}

We compare the measured angular 2PCF with the theoretical prediction derived for matter clustering to estimate quasar bias. The theoretical angular 2PCF is obtained from the angular power spectrum \(C_\ell\) via the Legendre transformation:
\begin{equation}
\omega_{\mathrm{m}}(\theta) = \frac{1}{4\pi} \sum_{\ell=\ell_{min}}^{\ell_{\mathrm{max}}} (2\ell + 1) C_\ell P_\ell(\cos \theta)\;,
\label{eq:matter_corr}
\end{equation}
where \(P_\ell(\cos \theta)\) is the Legendre polynomial of degree \(\ell\), and the multipole moment $\ell$ relates to the angular scale $\theta$ via $\ell=180^\circ/\theta$. The sum in the above equation should formally run from $\ell=1,...,+\infty$, but we restrict the summation to \(\ell_{\mathrm{min}} = 30\) and \(\ell_{\mathrm{max}} = 10^4\) since our angular binning ranges from 0.05\(^\circ\) to 1\(^\circ\).

{The angular power spectrum (PS) is derived by projecting the three-dimensional PS, $P(k,z)$, along the line of sight, typically using the Limber approximation:
\begin{equation}
\label{equ:ang_cl}
C_\ell = \int \frac{\mathrm{d}z}{H(z) \chi^2(z)} \left(\frac{\mathrm{d}N}{\mathrm{d}z}\right)^2P_{\text{m}}\left(\frac{\ell + 0.5}{\chi(z)}, z \right),
\end{equation}
where \(H(z)\) is the Hubble parameter and \(\chi(z)\) is the comoving radial distance to redshift \(z\). The function \(\mathrm{d}N/\mathrm{d}z\) represents the normalized redshift distribution of the quasar population for a given redshift bin; ideally this should be the {true} redshift distribution, which we do not know exactly not having spec-$z$s for the full sample. As discussed below, we will compare two approaches to approximate $dN/dz$, one by using photo-$z$s directly and the other by cross-matching with spec-$z$ samples.}

{The term \(P_{\mathrm{m}}(k, z)\) is the three-dimensional non-linear matter PS evaluated at the wavenumber \(k = \frac{\ell + 0.5}{\chi(z)}\) and redshift \(z\). To obtain it, for the matter-only case, we use the Core Cosmology Library (\texttt{CCL}; \cite{Chisari2018}) which uses the \textsc{CLASS} algorithm \citep{Blas2011} for the linear PS
and non-linear corrections to the matter PS are applied using the Halofit prescription
\citep{Smith2003, Takahashi2012}. This requires cosmological parameters and here we adopt the flat $\Lambda$CDM  cosmology with  $\Omega_c=0.2642$ (cold dark matter fraction), $\Omega_b=0.0493$ (baryonic matter fraction), $n_s=0.965$ (spectral index), $H_0=67.4  \mathrm{kms^{-1}Mpc^{-1}}$ (Hubble constant), $\sigma_8=0.811$ (amplitude of matter density fluctuations on a scale of 8 $h^{-1}$ Mpc) and a cosmological constant \citep{Planck2020}. In principle, the cosmological model could be varied in the analysis, however our measurements are not sufficiently sensitive to constrain both quasar bias and cosmological parameters, we therefore fix the latter to best-fit Planck values. The resulting $\omega_\mathrm{m}(\theta)$ curves, computed for each redshift bin using $dN/dz_{\mathrm{phot}}$ as the assumed redshift distribution, are shown as black dashed lines in Fig.~\ref{fig:correlation}. In Sec.~\ref{sec:bias} we detail how we use these theoretical predictions to derive quasar bias by comparing with the measured 2PCF. First we however discuss in Sec.~\ref{sec:covariance} the details of the covariance matrix needed for such calculations.}
}

\subsection{Covariance estimation}
\label{sec:covariance}

In order to derive quasar bias by comparing the observed 2PCF with its theoretical counterpart, we need the covariance matrix (CM) quantifying the correlations of $\omega(\theta)$ at different scales, which in our case are particular bins in $\theta$. We estimate the CM directly from the data via the {the Bootstrap method \citep{Mohammad2022}. It is a non-parametric internal error estimation method for a measured quantity, where multiple realizations of the dataset are generated by randomly sampling with replacement. In our analysis, we divided the KiDS DR4 survey footprint into $N_{\mathrm {patch}}=200$ spatial patches, and each bootstrap sample is constructed by randomly selecting $N_{\mathrm {patch}}$ patches with replacement, allowing some patches to be repeated while others may be omitted.}

For each bootstrap realization, $\omega(\theta)$ is recalculated using the selected patches. This procedure is repeated $N_{\mathrm {boot}}=15,000$ times, resulting in $N_{\mathrm {boot}}$ bootstrap realizations of $\omega(\theta)$. $N_{\mathrm{ boot}}$ is determined by checking that the eigenvalues of the covariance matrix have converged.
The bootstrap estimate of the covariance matrix is then computed as:

\begin{equation}
C_{ij} = \frac{1}{N_{\mathrm {boot}}-1} \sum_{k=1}^{N_{\mathrm{ boot}}} 
\left[ \omega_k(\theta_i) - \bar{\omega}(\theta_i) \right] 
\left[ \omega_k(\theta_j) - \bar{\omega}(\theta_j) \right],
\end{equation}
where $\omega_k(\theta_i)$ and $\omega_k(\theta_j)$ are the angular 2PCF measured in the $k$-th bootstrap sample at angular bins $\theta_i$ and $\theta_j$, respectively, and $\bar{\omega}(\theta_i)$ is the mean value across all bootstrap samples:

\begin{equation}
\bar{\omega}(\theta_i) = \frac{1}{N_{\mathrm{ boot}}} \sum_{k=1}^{N_{\mathrm{ boot}}} \omega_k(\theta_i).
\end{equation}

The inverse of the CM, $C_{ij}^{-1,boot}$, estimated using Bootstrap method is biased because of the finite number of subsamples; that is, $C_{ij}^{-1,boot}$ does not exactly correspond to the true inverse CM. We applied the Anderson-Hartlap-Kaufman (AHK) debiasing factor \citep{Hartlap2007,kaufman2008,Vakili2023} to correct for this bias in the inverse CM calculation.  AHK debiasing factor is defined as:
\begin{equation}
f_{\text{AHK}} = \frac{N_{\text{patch}} - N_{{d}} - 2}{N_{\text{patch}} - 1},
\end{equation}
where \(N_d=9\) is the number of angular bins used. 
The corrected inverse CM thus becomes:
\begin{equation}
\label{eq:covariance}
C_{\text{ij}}^{-1,AHK} = f_{\text{AHK}} \times C_{ij}^{-1,boot}.
\end{equation}
This corrected inverse covariance is then used for uncertainty-weighted model fitting to derive the quasar bias.
We use the $\tt{TreeCorr}$ package to compute the {Bootstrap} estimate of the CM. 
{$\tt{TreeCorr}$ employs k-means clustering using the angular positions (right ascension, declination) of quasars 
to generate the patches for covariance estimation. }

\begin{figure*}[!htbp] 
    \centering
    \includegraphics[width=1\textwidth]{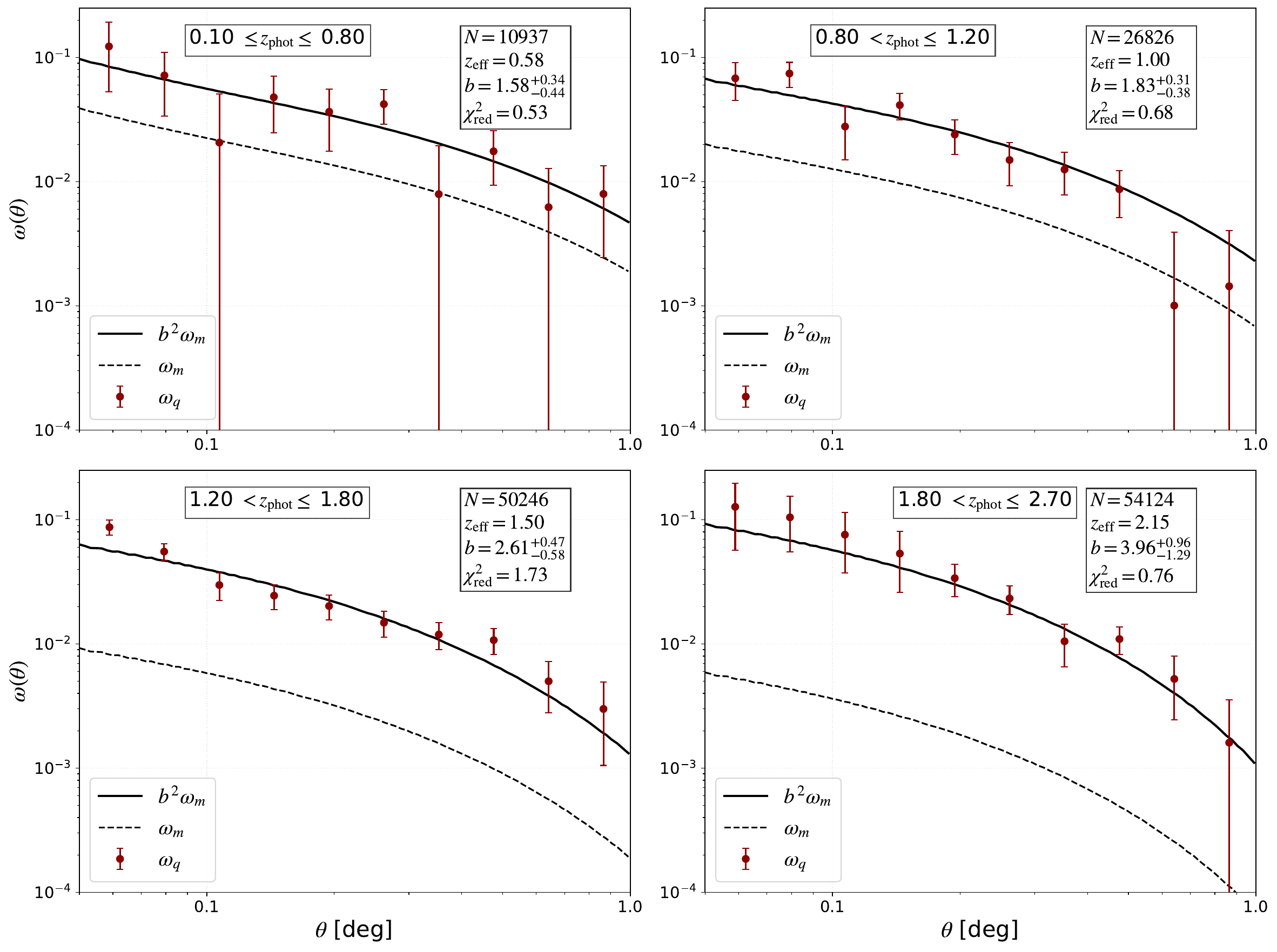}
    \caption{Angular auto-correlation function of KiDS DR4 quasars measured in four photometric redshift bins. Red data points with error bars are the quasar 2PCF measurements, while the dashed lines show the predicted matter 2PCF $\omega_m(\theta)$, obtained for Planck $\Lambda$CDM cosmology and quasar redshift distribution taken as $dN/dz_\mathrm{phot}$. The solid black line shows $b^2 \omega_m$, with the best-fit bias value indicated in the legend. $N$ and $z_{\rm eff}$ are the number of quasar objects and the effective redshift, respectively. The measurements shown here were corrected for stellar contamination using $p_\mathrm{star}=0.99$ as discussed in Sec.~\ref{sec:stellar_correction}.
    }
       
    \label{fig:correlation}
\end{figure*}

\subsection{Quasar bias and host halo mass estimation}
\label{sec:bias}

Here we detail how we derive the quasar effective bias of KiDS quasars, $b(z)$, from the angular correlation measurements.  Under the assumption of scale-independent biasing, the observed angular 2PCF of quasars is modeled as a scaled version of the matter $\omega(\theta)$, with the scaling governed by the square of the effective bias parameter: 
\begin{equation}
    \omega_{q}(\theta)=b^2(z) \omega_\mathrm{m}(\theta)\;.
\end{equation}
In each redshift bin, square of quasar bias \( b^2(z) \) is inferred through a chi-squared minimization:
\begin{equation}
\chi^2 = \sum_{i,j} \left[{\omega}_{q}({\theta_i}) - b^2(z) {\omega}_\mathrm{m}(\theta_i)\right]^{\mathrm{T}} 
{C}^{-1,AHK}_{ij}
\left[{\omega}_{q}({\theta_j}) - b^2(z) {\omega}_\mathrm{m}(\theta_j)\right],
\label{eq:chi2_bias_fit}
\end{equation}
In this analysis, we fit the observed quasar clustering over the angular range $0.05 \leq \theta (\mathrm{deg}) \leq 1.0$.
We employed a Monte Carlo approach to robustly estimate the goodness-of-fit \citep{Fumagalli2022}. We utilized the $\chi^2$ calibration method to account for covariance matrix estimation effects and correlated data points. This is done through moment-matching to a scaled $\chi^2$ distribution. The result is a calibrated reduced $\chi^2$, denoted as $\chi^2_{\rm red}$ in Fig.~\ref{fig:correlation}. We determine \(b^2\) in each redshift bin by minimizing the \(\chi^2\) statistic, and derive \(b\) and its uncertainties from the resulting \(\chi^2(b)\) curve. Owing to the asymmetry of \(\chi^2(b)\), the inferred bias values have asymmetric 
\(1\sigma\) uncertainties. These are presented in Figs.~\ref{fig:correlation} and \ref{fig:impact_of_redshift_distribution}, as well as in Table.~\ref{table:bias}.

} 

Given a normalized redshift distribution 
\(dN/dz\), the effective redshift in each bin is computed as:
\begin{equation}
z_{\mathrm{eff}} = \frac{\int_{z_{\mathrm{min}}}^{z_{\mathrm{max}}} z \left( \frac{\mathrm{d}N}{\mathrm{d}z} \right) \mathrm{d}z}{\int_{z_{\mathrm{min}}}^{z_{\mathrm{max}}} \left( \frac{\mathrm{d}N}{\mathrm{d}z} \right) \mathrm{d}z}
\label{eq:zeff}
\end{equation}
In Fig.~\ref{fig:correlation}, the $z_\mathrm{eff}$ was computed using photometric
$dN/dz$, which we take as our fiducial approach.

The fitted quasar bias increases steadily from $\sim1.6$ in the first redshift bin 
to $\sim4.0$ in the last $z$-bin (Fig.~\ref{fig:correlation} and Table \ref{table:bias}). 
This trend agrees with the expectations from halo bias models and previous studies \citep{Myers2007a,Ross2009,Eftekharzadeh2019}. 
{ 
We calculated the effective peak height $\nu_{\mathrm {eff}}$ directly from our bias measurements using the \cite{Tinker2010} and \cite{Comparat2017} (hereafter \citetalias{Tinker2010} and \citetalias{Comparat2017} respectively) bias–$\nu_{\mathrm {eff}}$ relations, and derived the effective host halo masses $M_{\mathrm {eff}}$ for each redshift bin by inverting their halo mass–bias relations.
Table~\ref{table:halo_mass} presents the $\nu_{\mathrm {eff}}$ and \(\log_{10}M_{\mathrm {eff}}(h^{-1}M_\odot)\) 
and their lower and upper bounds at each effective redshift. 
For both models, \(\nu_{\mathrm{eff}}\) increases with redshift, and the
corresponding halo masses lie in the range
\(\log_{10} M_{\mathrm{eff}} (h^{-1} M_\odot) \simeq 12.7\text{--}12.9\). 
These trends are consistent with the widely accepted picture of quasar host halo mass \citep{Shen2009, Eftekharzadeh2019, Petter2023, Pizzati2024, Giner2025}. 
We note that the selection effects of our quasar sample, particularly at high redshifts, are not accounted for in this analysis.} {Table~\ref{table:halo_mass_pastworks} presents a comparison of quasar host halo 
mass estimates from the literature alongside our results based on \citetalias{Comparat2017}. A direct 
comparison across studies requires caution for several reasons, 
including differences in the adopted halo mass statistic, cosmological parameters, redshift ranges, and sample selection criteria such as luminosity thresholds, photometric versus spectroscopic identification, and survey depth. \cite{Giner2025} and \cite{Eftekharzadeh2019} report {minimum halo masses}; \cite{Pizzati2024} reports an {average halo mass}; while \cite{Petter2023}, \cite{Shen2009}, and this work report {effective halo masses}. Nevertheless, we find that our estimates remain consistent with the literature, and the typical values are $\sim10^{12.5}-10^{13} h^{-1} M_\odot$.}

\subsection{Stellar contamination correction}
\label{sec:stellar_correction}

Stellar contamination of the quasar sample is a possibly significant factor affecting the clustering measurements and quasar bias derivation. Stars are expected to have a nearly flat angular 2PCF, in contrast to steep power-law for extragalactic sources such as quasars. Stellar contamination would therefore affect the slope of the measured quasar 2PCF by making it less steep.

In order to inspect and mitigate the stellar contamination effect, we take a similar approach as in \cite{Myers2006}. Namely, we model the observed angular 2PCF of our sample as being composed of a true quasar clustering component and that from stars, appropriately weighted by the purity of the quasar sample.\footnote{{In what follows we assume that only stars can contaminate the quasar sample, i.e. we neglect possible contamination from non-QSO galaxies.}} As discussed in \cite{Myers2006}, it can be shown that this leads to the following expression for true quasar clustering (neglecting cross-terms from star-quasar correlations):

\begin{equation}
\omega_{\mathrm{q}}(\theta) = \frac{\omega_{\mathrm{obs}}(\theta) - (1 - a^2)  \omega_{\mathrm{{s}}}(\theta)}{a^2}\;,
\label{eq:stellar_correction}
\end{equation}
where $\omega_\mathrm{s}(\theta)$ is angular 2PCF of stars, while $a$ is quasar sample purity, for which we take $a=0.98$ from   
the KiDS DR4 QSO classification model of \citetalias{Nakoneczny2021}. 

We measure the star 2PCF $\omega_s$ from samples of star candidates provided by the \citetalias{Nakoneczny2021} classification model
\footnote{\url{https://kids.strw.leidenuniv.nl/DR4/quasarcatalog.php}}. {For that, we select}  
objects based on different stellar classification probability thresholds, \(p_{\mathrm{star}}\). A threshold of \(p_{\mathrm{star}} \geq 0.99\) yields a star catalog of higher purity, containing fewer quasars misclassified as stars compared to catalogs with \(p_{\mathrm{star}} \geq 0.90\) and \(p_{\mathrm{star}} \geq 0.75\). The autocorrelation function of these stellar samples, \(\omega_{\mathrm{s}}(\theta)\), was measured over the angular range \(0.05^\circ \leq \theta \leq 1.0^\circ\)  following the procedure described in Sec.~\ref{sec:observed}, and the results are presented in Fig.~\ref{fig:stellar_correlation}. 
Notably, the clustering amplitude of the \(p_{\mathrm{star}} \geq 0.99\) sample remains approximately flat across the considered angular scales, indicating a low residual clustering signal from extragalactic sources, while
the angular 2PCF for the \(p_{\mathrm{star}} \geq 0.90\) and \(p_{\mathrm{star}} \geq 0.75\) samples exhibit a clear decline with increasing angular separation.

We used the above-discussed measurements of $\omega_s(\theta)$ for the various $p_\mathrm{star}$ thresholds in Eq.~\eqref{eq:stellar_correction} to correct the quasar clustering measurements for stellar contamination. These corrected measurements were further employed to estimate quasar bias as discussed in Sec.~\ref{sec:bias}.
We found no significant variation in the bias values  
across the different \(p_{\mathrm{star}}\) thresholds. This suggests that our bias estimates are robust to moderate levels of residual stellar contamination.
For the final analysis and interpretation, we adopted the stellar catalog with \(p_{\mathrm{star}} \geq 0.99\) for the stellar contamination correction in the quasar clustering measurements
(Fig.~\ref{fig:correlation}).

\begin{table}
\centering
\caption{Effective peak height $\nu_{\mathrm{ eff}}$ and halo mass, \(M_{\mathrm {eff}}\) of KiDS DR4 quasars obtained by using \citetalias{Tinker2010} and \citetalias{Comparat2017} models for each redshift $z_{\mathrm {eff}}$. }
\label{table:halo_mass}

\begingroup
\renewcommand{\arraystretch}{1.5}

\begin{tabular}{lccc}
\hline\hline
Model & $z_{\mathrm {eff}}$ & $\nu_{\mathrm {eff}}$ & $\log_{10}M_{\mathrm {eff}} (h^{-1}M_\odot)$ \\
\hline\hline

\citetalias{Tinker2010} 
& 0.58 & $1.54^{+0.25}_{-0.41}$ & $12.86^{+0.18}_{-0.22}$ \\
& 1.00 & $1.73^{+0.21}_{-0.29}$ & $12.65^{+0.11}_{-0.13}$ \\
& 1.50 & $2.21^{+0.24}_{-0.34}$ & $12.70^{+0.07}_{-0.07}$ \\
& 2.15 & $2.85^{+0.38}_{-0.61}$ & $12.73^{+0.13}_{-0.15}$ \\

\hline

\citetalias{Comparat2017}
& 0.58 & $1.54^{+0.25}_{-0.41}$ & $12.93^{+0.17}_{-0.21}$ \\
& 1.00 & $1.73^{+0.21}_{-0.29}$ & $12.68^{+0.12}_{-0.13}$ \\
& 1.50 & $2.21^{+0.24}_{-0.34}$ & $12.72^{+0.07}_{-0.07}$ \\
& 2.15 & $2.85^{+0.38}_{-0.61}$ & $12.76^{+0.13}_{-0.15}$ \\

\hline\hline
\end{tabular}
\endgroup
\end{table}

\begin{table}
\caption{{Comparison of quasar host halo mass estimates between this work and literature.}
}

\label{table:halo_mass_pastworks}
\begin{threeparttable}

\begin{tabular}{ccc}
\hline
{Work }& ${z_{\rm eff}}$ & $\log_{10}M\,(h^{-1}\,M_\odot)$ \\

\hline

&0.58&$12.93^{+0.17}_{-0.21}$\\
This work (using \citetalias{Comparat2017} model)\tnote{3}&1.00&$12.68^{+0.12}_{-0.13}$\\
&1.50&$12.72^{+0.07}_{-0.07}$\\
&2.15& $12.76^{+0.13}_{-0.15}$\\
\hline

\multirow{4}{*}{\cite{Giner2025} \tnote{1}} 
& 0.5 &13.03\\  
& 1.5 & 12.91\\ 
& 2.5 & 12.96\\ 
& 3.5 & 12.99\\
\hline

\cite{Pizzati2024} \tnote{2} 
& 6.0 & 12.33\\
\hline

\cite{Petter2023}\tnote{3}\\
Unobscured QSO 
& 1.4 & 12.58
\\
 Obscured QSO  
& 1.4 & 12.93
\\
\hline

\cite{Eftekharzadeh2019} \tnote{1} 
& 1.5 & 12.36\\
\hline

\cite{Shen2009}\tnote{3} & & \\
Radio-loud QSO & 1.5 & $\sim 13.0$  
\\
Radio-quiet QSO & 1.5 & $\sim 12.30$
\\

\hline

\end{tabular}
\begin{tablenotes}
\footnotesize
\item[1] Minimum halo mass
\item[2] Average halo mass
\item[3] Effective halo mass 
\end{tablenotes}
\end{threeparttable}
\end{table}

\subsection{Impact of the assumed redshift distribution}
\label{sec:impact_of_redshift_distribution}
\begin{figure}[!htbp] 
    \centering
    \includegraphics[width=\columnwidth]{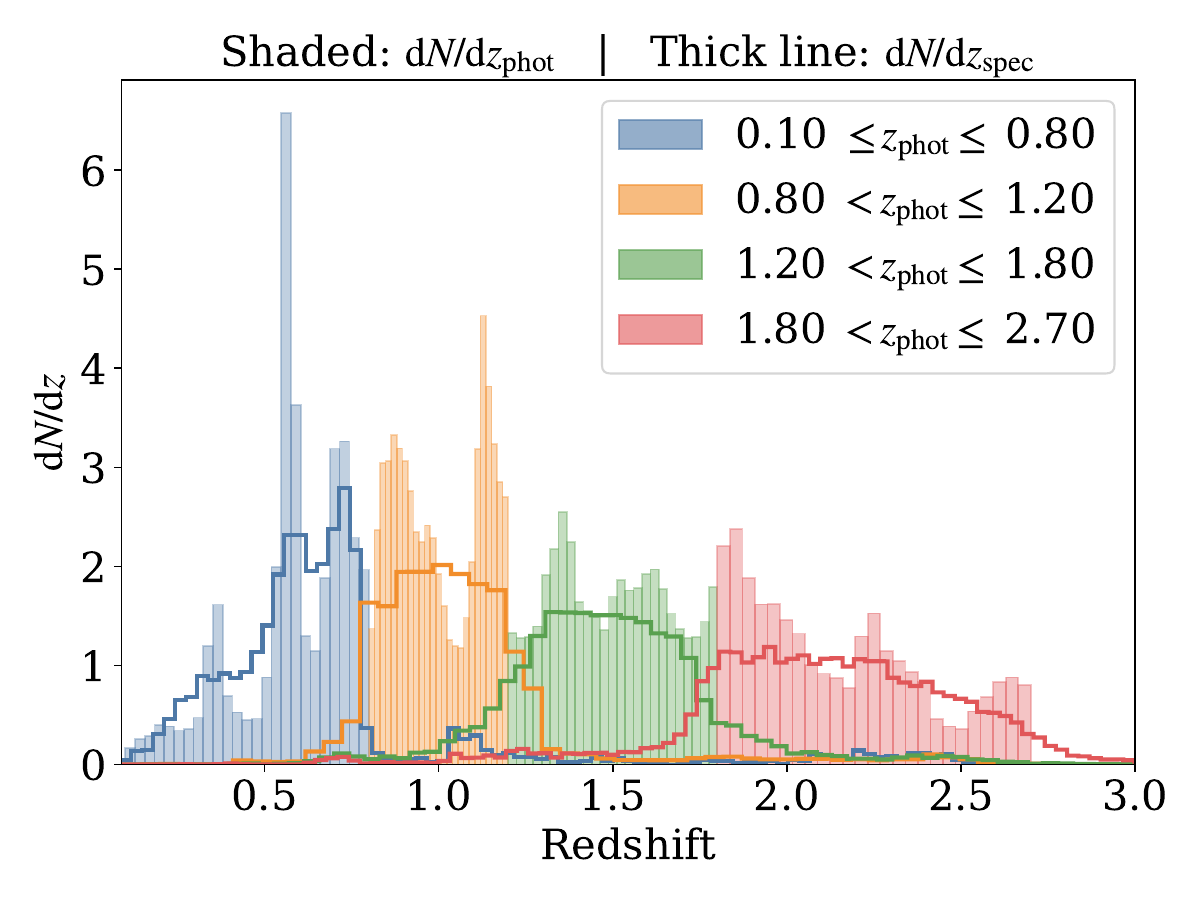}
    \caption{
Quasar redshift distributions for the four tomographic bins used in this study.
Shaded histograms are the photo-$z$
distributions of the KiDS DR4 QSO sample, while the thick lines correspond to the spec-$z$  
distributions obtained by direct cross-match of the quasars from a given photo-$z$ bin with overlapping DESI and SDSS spectroscopy. Each distribution is normalized to unity under the curve.
}

    \label{fig:nof_z}
\end{figure}

Bias inference from the angular 2PCF depends on the assumed redshift distribution $dN/dz$, as {this quantifies the projection of the 3D quasar distribution onto the sky, and} enters the model quadratically 
(see Eq.~\ref{equ:ang_cl}).
To investigate the impact of {the assumed} redshift distribution on quasar bias measurements, we {compared two approaches. The fiducial one, for which the results are presented in Fig.~\ref{fig:correlation}, uses the photometric estimates, $dN/dz_\mathrm{phot}$, directly. The second method follows \cite{Myers2007a}: KiDS DR4 quasars in each of the photo-$z$ bins are cross-matched with the DESI+SDSS spec-$z$ quasar sample, which gives $dN/dz_\mathrm{spec}$ for a given bin.}
The derived $dN/dz$ {are compared}  
in Fig.~\ref{fig:nof_z},{where shaded bars present the photo-$z$ distributions, while the} thick lines {illustrate $dN/dz_\mathrm{spec}$. In that plot, each of the tomographic redshift distributions is normalized to unity under the curve to appreciate better the differences in the redshift support for each of the bins.}

{By design, the $dN/dz_\mathrm{phot}$ are truncated at bin edges and do not reflect the photo-$z$ uncertainties scattering the objects between the bins. The true redshift distributions will therefore always be broader than the photo-$z$s suggest. What is more, at higher redshifts, photo-$z$ errors generally increase, leading to a larger fraction of  outliers. Consequently, \( dN/dz_{\mathrm{phot}} \) will deviate more substantially from the true underlying redshift distribution as the redshifts increase.}

On the other hand, the spec-$z$ distributions derived via the cross-match are broader than the photometric ones and indicate some outliers at values far apart from bin centers. We should however
emphasize that the spec-$z$ distribution derived from the cross-matched sample is not necessarily representative of the true redshift distribution of our entire quasar sample, as the \citetalias{Nakoneczny2021} selection may contain quasar candidates not represented by DESI or SDSS in the color space.
Therefore, bias values inferred using \( {dN}/{dz}_{\mathrm{spec}}\) may not be inherently more accurate than those obtained from the photo-$z$ distribution of the entire sample.

Both the shape and width of the redshift distribution significantly influence the theoretical predictions of the angular 2PCF. If \( dN/dz \) is narrower, the integrand in Eq.~\eqref{equ:ang_cl} results in a {higher angular} 
clustering amplitude compared to that of a broader distribution. This in turn will lead to {lower} best-fit quasar bias for the narrower $dN/dz$ for a given redshift bin. This is indeed what we find, as illustrated in Fig.\ref{fig:impact_of_redshift_distribution} and quantified in Table.\ref{table:bias}. Both compare the best-fit effective bias derivations for the KiDS DR4 quasars when using $dN/dz_\mathrm{phot}$ (blue points in Fig.\ref{fig:impact_of_redshift_distribution}) or $dN/dz_\mathrm{spec}$ (green points) in the modeling.{ In Fig.\ref{fig:impact_of_redshift_distribution} we additionally show simple phenomenological fits for the the redshift dependence of quasar bias:   $b(z) = b_0 + b_1 z + b_2 z^2$ for $dN/dz_\mathrm{phot}$ and $dN/dz_\mathrm{spec}$. The best-fit values for $b_0, b_1, b_2$ are provided in the plot while their errors are listed in Table \ref{tab:quad_fit} and illustrated as shaded bands in the Figure. We note that such quadratic relations fit the datapoints very well. On the other hand, as we do not account for redshift-bin correlations (which would require building a relevant covariance matrix), the fit uncertainties will be somewhat underestimated. We can however firmly conclude that}
using spectroscopic redshifts to model $dN/dz$ gives bias values larger than the photometric case for the considered redshift range.

These results show that the choice of the underlying redshift distribution has a non-negligible impact on the inferred quasar bias estimation for photometric samples. 
This emphasizes the importance of reliable $dN/dz$ calibration for clustering-based cosmological analyses with such quasar datasets. 
Here, we adopt the quasar bias derived from the \( dN/dz_{\mathrm{phot}} \) of the full sample as our fiducial result, because this redshift distribution corresponds directly to the dataset used in the observed angular 2PCF measurements, unlike the spec-$z$ distribution of the cross-match, which may not be fully representative of the entire sample.

\begin{figure}[!htbp] 
    \centering
    \includegraphics[width=0.5\textwidth]{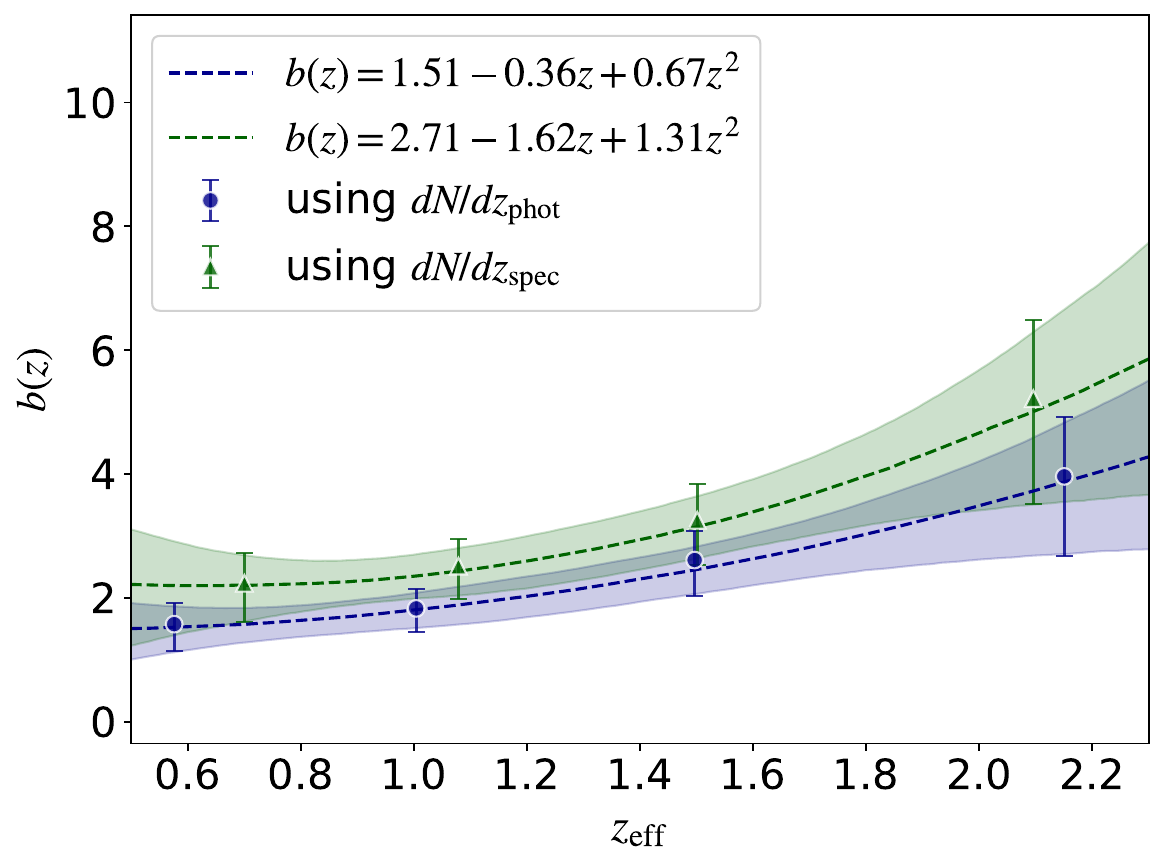}
        \caption{Effective bias of KiDS DR4 quasars as a function of redshift,  estimated from angular clustering analysis. {We present results from two modeling choices:} 
        blue circles {were obtained by assuming $dN/dz_{\mathrm{phot}}$ directly as the underlying redshift distribution in the photo-$z$ bins, while} for green triangles we used per-bin cross-matches of KiDS photometric quasars with DESI and SDSS spectroscopy. Other model assumptions are the same in both cases.
        The dashed lines indicate the best-fit models of the form $b(z) = b_0 + b_1 z + b_2 z^2$. The corresponding best-fit parameters are shown in the legend, while their uncertainties are reported in Table~\ref{tab:quad_fit}. The shaded regions represent the $\pm1\sigma$ uncertainty bands of the fitted models.}
        \label{fig:impact_of_redshift_distribution}
    \end{figure}

\subsection{Comparison with previous derivations of quasar bias}
\label{sec:bias_comparison}
The effective bias of quasars has been measured in a number of past studies and here we compare our results with the literature. For this comparison, we chose those where the large-scale bias was estimated in tomographic bins, both using photometric and spectroscopic redshifts. These are in particular: \cite{Croom2005}, who
{employed the spectroscopic sample from the 2dF QSO Redshift Survey \citep[2QZ,][]{Croom2004}  covering the  range $0.3<z<2.2$};
already discussed 
\cite{Myers2007a}, who used {photometrically
classified quasars drawn from SDSS DR4 spanning \(0.4<z< 2.8\)}; \cite{Ross2009}, who {presented bias measurements for spectroscopic quasars from SDSS DR5 within the redshift range \(0.3\leq z\leq2.2\)}; and \cite{Laurent2017}, who {studied eBOSS spectroscopic quasar sample for redshifts \(0.9<z<2.2\)}. We note that these past works used different cosmological parameters than us, care should be therefore taken when comparing the bias derivations. The main factor influencing the bias estimate will be different adopted $\sigma_8$, which rescales the amplitude of the theoretical power spectrum, and hence directly affects the best-fit  effective bias. Moreover, the previous
analyses reported $b(z)$ values at the mean redshifts of the adopted bins,
which depends on the assumed redshift distributions per bin.

We illustrate the comparison of our results from this paper with the past quasar bias measurements in Fig.~\ref{fig:bias_comparison_past}.  We observe very good consistency of our estimates with the 2QZ and SDSS derivations respectively by \cite{Croom2005} and \cite{Myers2007a,Ross2009}, while the more recent eBOSS constraints by \cite{Laurent2017} lie systematically below ours, especially for the highest-$z$ bins. This difference is enhanced especially due to considerably smaller errorbars on $b(z)$ in this latter paper as compared to other works, including ours. However, the discrepancy hardly exceeds $2\sigma$ even for the most diverging datapoint at $z\sim2$, which is appreciated especially when compared to our best-fit quadratic model of the form $b(z)=1.51-0.36 z +0.67 z^2$ (Sec.~\ref{sec:bias} \& Fig.~\ref{fig:impact_of_redshift_distribution}). This overall consistency between our quasar bias fits and those from the previous analyses is especially remarkable taking into account various surveys and sample selections, as well as cosmologies used.
We conclude that within the explored redshift ranges and scales the bias measurement for quasars is relatively insensitive to such factors.

{As an additional diagnostic, we also examined a full-redshift angular correlation measurement and a fixed-slope power-law/Limber inversion in order to compare more directly with broad-redshift analyses such as \cite{Petter2023} and \cite{Eltvedt2024}. These checks confirm that such comparisons are difficult to interpret quantitatively for our sample: the redshift distributions, selections, effective redshifts, and available covariance information differ between studies, and the full-redshift angular correlation function is not well described by a single power law. We therefore retain the tomographic bias measurements as our fiducial result.}

\begin{figure}[!htbp] 
    \centering
    \includegraphics[width=0.5\textwidth]{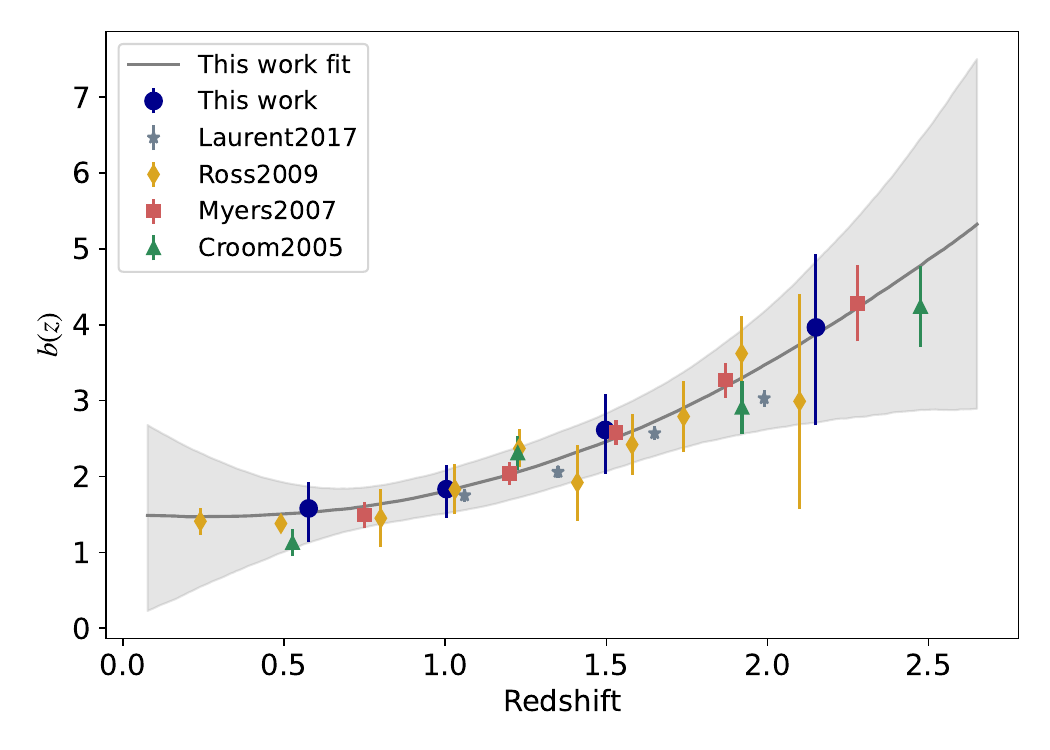}
    \caption{{Comparison of the the effective quasar bias
    as a function of redshift {between this work and previous analyses.} 
    The dark blue circles represent the KiDS DR4 measurements from this paper, while the others are respectively: 2QZ from \cite{Croom2005} (upward green triangles); SDSS DR4 from \cite{Myers2007a} (red squares); SDSS DR5 from \cite{Ross2009} (orange diamonds); and eBOSS from \cite{Laurent2017} (downward grey triangles). The background grey line with error band is our best-fit quadratic $b(z)$.}}
    \label{fig:bias_comparison_past}
\end{figure}

\section{Conclusions}
\label{sec:conclusion}

This study provides the first comprehensive measurement of angular clustering and scale-independent {effective} bias for photometrically classified 
quasars {from the Kilo-Degree Survey Data Release 4} \citep{Kuijken2019,Nakoneczny2021}. {Using the original quasar selection from this latter paper \citepalias{Nakoneczny2021}, we restricted our analysis to the safe sample,} 
where object features lie within the range of the training data of the QSO classification model.
Although we used the {previous} photometrically classified quasar catalog, we did not adopt the 
photometric redshifts derived {originally} by \citetalias{Nakoneczny2021}. Instead, we applied our new deep learning model, Hybrid-z, {developed earlier by \cite{Anjitha2025}} 
Hybrid-z combines 
KiDS \(\mathrm{ugri}\) imaging  with KiDS+VIKING nine-band photometry (\(\mathrm{ugriZYJHK_s}\)) and {in this study it} is trained on DESI DR1 and SDSS DR17 spectroscopic {quasars}. This {way we update the} approach 
of \citetalias{Nakoneczny2021}, who relied solely on SDSS DR14 and used only colors and magnitudes without incorporating imaging information. 

Our new model significantly improves the previous KiDS QSO photo-$z$ derivation in practically all the statistics, including the mean bias and scatter of residuals.
On the other hand, despite better statistical accuracy and precision of our photo-$z$s, we still observe the characteristic artefacts in quasar redshift estimates as in past studies: redshift focusing is due to the emission line shifts across photometric filters, resulting in narrow peaks in photometric redshift distributions.

Having tested the Hybrid-z model on spectroscopic data with known redshifts, we trained it on the overlap between KiDS DR4 and DESI+SDSS quasars, and applied it to the entire photometric sample of safe quasars from \citetalias{Nakoneczny2021}. The final product, used for the clustering analysis, is an updated KiDS DR4 QSO catalog containing about 157k objects on an effective area of $\sim777$ deg$^2$.

{We divided our
quasar sample into four {photometric} redshift bins in the 
range \(0.1\leq z \leq2.7\), where most of the objects are located.
For each of the bins, we measured the quasar angular {two-point correlation function} (2PCF) over scales from $0.05^\circ$ to $1^\circ$. 
These measurements were then compared with theoretical predictions for dark-matter 2PCF assuming \cite{Planck2020} $\Lambda$CDM cosmological parameters. Using a simple bias relation of the form $\omega_Q(\theta) = b^2(z) \omega_m(\theta)$, we obtained best-fit bias values per bin.
Our analysis of quasar clustering exhibits a clear increase in bias with redshift,  from $b \sim 1.6$ at $z \sim 0.6$ to $b \sim 4.0$ at $z \sim 2.2$. We calculated the effective peak height $\nu_{\mathrm{eff}}$ and host halo mass $M_{\mathrm {eff}}$ of our quasars.$\nu_{\mathrm{eff}}$ increases from $\sim 1.5$ at $z \sim 0.6$ to $\sim 2.85$ at $z\sim2.15$.
This trend is consistent with halo bias models and previous observational studies, confirming that quasars observed at higher redshifts reside in progressively more massive dark matter halos.

We have assessed the impact of stellar contamination on our quasar clustering measurements by applying the correction formula from \citet{Myers2006}, using angular scale-dependent stellar clustering based on stellar samples with varying classification probabilities. Our analysis shows that the inferred quasar bias remains largely unchanged across different stellar purity thresholds, indicating that the results are robust to moderate levels of residual stellar contamination. For the final analysis, we adopt the highest-purity stellar sample ($p_\mathrm{star} \geq 0.99$) to correct the quasar angular correlation function, ensuring reliable bias estimates.

{We also investigated {how our} quasar bias {derivations depend} on the per-bin redshift distributions employed in the theoretical model. We used two approaches: taking $dN/dz_\mathrm{phot}$ for each bin directly, or employing  as redshift distributions the cross-matches of the bins with spectroscopic quasars from DESI+SDSS.
These latter $dN/dz$ are broader than the photometric ones due to photo-$z$ errors, and display tails far away from bin centers related to
catastrophic outliers. 
Consequently, using $dN/dz_{\mathrm {spec}}$ yields systematically higher bias values compared to $dN/dz_{\mathrm{ phot}}$ { as the amplitudes of the theoretical predictions for dark matter 2PCF are lower in the former case than in the latter}. Nevertheless, the spectroscopic distribution is not fully representative of the entire sample, {especially that the original quasar selection from \citetalias{Nakoneczny2021} that we employ was trained on SDSS DR14.} 
We therefore adopt the bias derived from the photo-$z$ distribution of the full KiDS DR4 quasar sample as our fiducial result.
For the future robust clustering-based cosmological inference from photometric quasar samples, {such as from the final KiDS DR5 or forthcoming LSST,} accurate characterization of the underlying redshift distribution will be crucial.}

Our effective quasar bias measurements {for the four effective redshifts of the bins can be} modeled as $b(z) = b_0 + b_1 z + b_2 z^2$ with best-fit parameters $(b_0, b_1, b_2) = (1.51, -0.36, 0.67)$. {A comparison with the previous derivations of quasar $b_Q(z)$, using samples such as 2QZ
\citep{Croom2005}, SDSS: photometric \citep{Myers2007a} and spectroscopic \citep{Ross2009}, as well as eBOSS \citep{Laurent2017}, shows overall consistency with our results, except for the latter work where the highest-$z$ bias estimates depart from our best-fit by more than $2\sigma$. A caveat regarding these comparison is that these past works used different cosmologies than us, which is especially important when $\sigma_8$ are not matching, as these directly rescale the matter power spectrum.

In this work, we estimated {updated} photo-$z$s {for quasars selected from the previous KiDS DR4, and performed the first measurements of KiDS quasar clustering and of their} scale-independent effective bias. 
In the near future, we plan to extend such studies to the final KiDS DR5 \citep{Wright2024}. A quasar selection from that release was already presented in \cite{Feng2025}, where, however SDSS data were used to train the classifier, and updating that to DESI should allow us to further improve the previous classification, especially at the faint end. Other avenues to explore it to use all 9 KiDS+VIKING passbands in the convolutional part of the Hybrid-z model, i.e., incorporate both imaging and magnitudes from the whole $u$ to $K_s$ range.

Angular 2PCF measurements give high signal-to-noise down to $0.1^\circ$ (at least) which could be further explored with 
more sophisticated theoretical frameworks, such as the halo occupation distribution \citep{Mitra2018,Eftekharzadeh2019,Petter2023, Chowdhary2025}.
Furthermore, both the quasar bias and other cosmological parameters {could be also derived} by cross-correlating the quasar distribution with CMB {lensing} \citep{Hirata2008,
Sherwin2012,Eltvedt2024, deBelsunce2025}
A crucial ingredient for the robustness of the cosmological constraints when using photometric quasars will be to properly calibrate their redshift distributions, and techniques such as self-organizing maps could be employed for that purpose \citep{Jalan2024, Wright2025}}.

On a longer term, forthcoming big data from, e.g., the Vera Rubin Observatory Legacy Survey of Space and Time \citep[LSST,][]{Ivezic2019}, as well as from the 4-metre Multi-Object Spectroscopic Telescope \citep[4MOST,][]{deJong2019} will allow to build and explore new quasar samples over most of the sky. This gives great perspectives not only to understand better their large-scale clustering properties, but also to connect these to smaller-scale environments of the cosmic web. 

\section*{Code availability}
{The Deep learning framework for the photometric redshift estimation, named Hybrid-z, will be publicly available at \url{https://hybrid-z.readthedocs.io/en/latest/}.}

\begin{acknowledgements}
We thank Angus Wright for his valuable comments and suggestions on the manuscript, and Andrej Dvornik and Ziang Yan for their helpful feedback and discussions during the early stages of this project.
\\
This work is supported by the Polish National Science Center through grants no. 2020/38/E/ST9/00395, and {2020/39/B/ST9/03494}. We thank Polish Ministry of Science and Higher Education under the agreement no. 2026/WK/06 for the infrastructure support.\\
Based on data products from observations made with ESO Telescopes at the La Silla Paranal Observatory under program IDs 177.A-3016, 177.A-3017 and 177.A-3018, and on data products produced by Target/OmegaCEN, INAF-OACN, INAF-OAPD and the KiDS production team, on behalf of the KiDS consortium. OmegaCEN and the KiDS production team acknowledge support by NOVA and NWO-M grants. Members of INAF-OAPD and INAF-OACN also acknowledge the support from the Department of Physics \& Astronomy of the University of Padova, and of the Department of Physics of Univ. Federico II (Naples).\\
We have made use of \textsc{TOPCAT} \citep{Taylor2005} software, as well as of \textsc{python} (\url{www.python.org}), including the packages \textsc{NumPy} \citep{harris2020}, \textsc{SciPy} \citep{Virtanen2020}, {\textsc{colossus} \citep{diemer2017}}, and \textsc{Matplotlib} \citep{Hunter2007}.
\end{acknowledgements}

\bibliographystyle{mnras}
\bibliography{references}

\begin{appendix}

\FloatBarrier
\section{Measured bias values}
\begin{table}[htbp]
\caption{
Comparison of effective redshifts ($z_{\mathrm{eff}}$) and bias values obtained using photo-$z$ distributions of KiDS DR4 quasar sample (\( dN/dz_{\text{phot}} \)) and spec-$z$ distributions of KiDS DR4 $\times$ (DESI DR1 + SDSS DR17) quasar data (\( dN/dz_{\text{spec}} \)) for the theoretical prediction of angular clustering.}
    \centering
    \begin{tabular}{|c|c|c|c|c|}
        \hline
        \rule{0pt}{3.5ex}
        photo-$z$ bin
        & \multicolumn{2}{c|}{using \( dN/dz_{\text{phot}} \)} 
        & \multicolumn{2}{c|}{using \( dN/dz_{\text{spec}} \)}\\[2pt]
        \cline{2-5}
        \rule{0pt}{3.5ex}
        & \( z_{\text{eff}} \) & bias & \( z_{\text{eff}} \) & bias \\[4pt]
        \hline
        \rule{0pt}{3ex}
        \( 0.1 \leq z_{\text{phot}} \leq 0.8 \) & 0.58 & $1.58_{-0.44}^{+0.34}$ & 0.70 & $2.24_{-0.62}^{+0.48}$ \\[2pt]
        \(0.8<z_{\text{phot}}\leq 1.2\) &1.00  & $1.83_{-0.38}^{+0.31}$  & 1.08&$2.52_{-0.53}^{+0.43}$  \\[2pt]
        \(1.2<z_{\text{phot}}\leq 1.8\) & 1.50 & $2.61_{-0.58}^{+0.47}$  & 1.50 & $3.26_{-0.72}^{+0.59}$  \\[2pt]
        \(1.8<z_{\text{phot}}\leq 2.7\) &2.15  &$3.96_{-1.29}^{+0.96}$  & 2.10 & $5.22_{-1.69}^{+1.27}$ \\[2pt]
        \hline
    \end{tabular}
    \label{table:bias}
\end{table}

\FloatBarrier
\section{Angular clustering of stars}
\begin{figure}[htbp]
    \centering
        \includegraphics[width=0.5\textwidth]{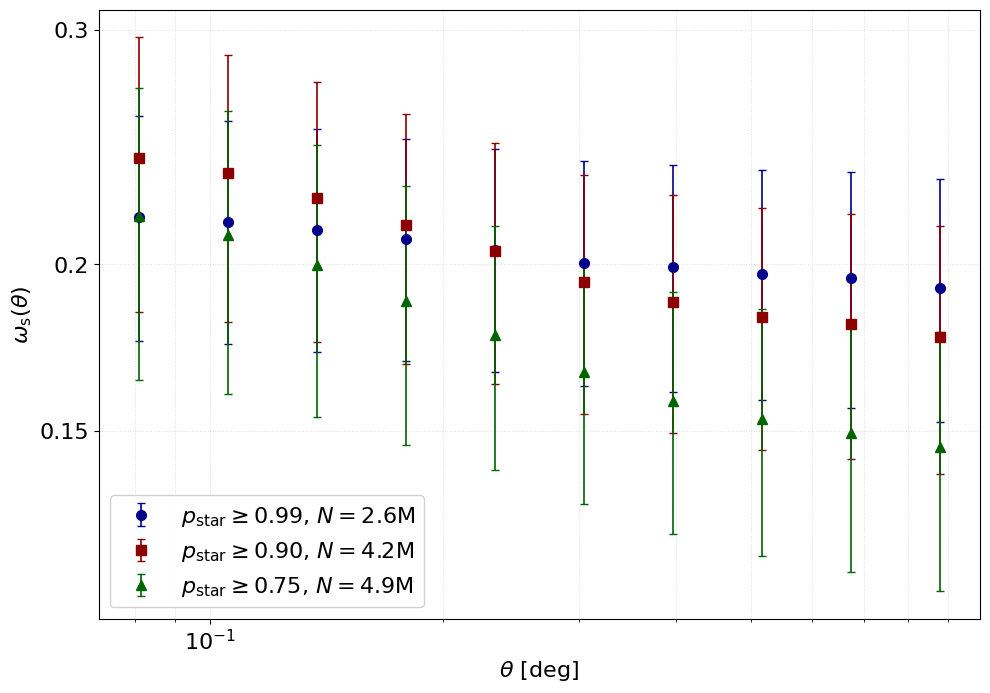}
        \caption{
        Angular two-point correlation function for stellar sources, $\omega_{\mathrm{s}}(\theta)$, measured from 0.05 to 1.0 deg selected by different stellar classification probability ($p_{\mathrm{star}}$) threshold. The corresponding number of sources \(N\) is indicated in the legend.}
        \label{fig:stellar_correlation}
\end{figure}

\FloatBarrier
\section{Best-fit parameters}
\label{sec:quad_fit}

Best-fit parameters and corresponding covariance matrix for the quadratic fit,$b_0+b_1z+b_2z^2$, describing the evolution of quasar bias with redshift as follows. Covariance matrix of best-fit parameters when using $dN/dz_{\rm phot}$ is:
\[\rm{C_{ij}}=
\begin{bmatrix}
 2.06 & -3.65 &  1.41 \\
-3.65 &  6.84 & -2.75 \\
 1.41 & -2.75 &  1.16
\end{bmatrix}
\]
and using $dN/dz_{\rm spec}$ is:
\[\rm{C_{ij}}=
\begin{bmatrix}
  7.38 & -12.16 &   4.55 \\
-12.16 &  20.9  &  -8.1  \\
  4.55 &  -8.1  &   3.26
\end{bmatrix}
\]

\begin{table}[!t]
\centering
\caption{The parameter uncertainties, corresponding to $1\sigma$ deviation, are derived from the square root of the diagonal elements of their covariance matrix.}
\label{tab:quad_fit}
\begin{tabular}{|c|c|c|}
    \hline
    Parameters & Using \(dN/dz_{\text{phot}}\) & Using \(dN/dz_{\text{spec}}\) \\[1pt]
    \hline
    \(b_0\) & \(1.51 \pm 1.44\) & \(2.71 \pm 2.72\) \\[1pt]
    \(b_1\) & \(-0.36 \pm 2.62\) & \(-1.62\pm 4.57\) \\[1pt]
    \(b_2\) & \(0.67 \pm 1.08\) & \(1.31 \pm 1.81\) \\[1pt]
    \hline
\end{tabular}
\end{table}

\section{Comparison of clustering measurements with previous studies}
\label{clustering_comparison}
\begin{figure}
    \centering
    \includegraphics[width=0.9\linewidth]{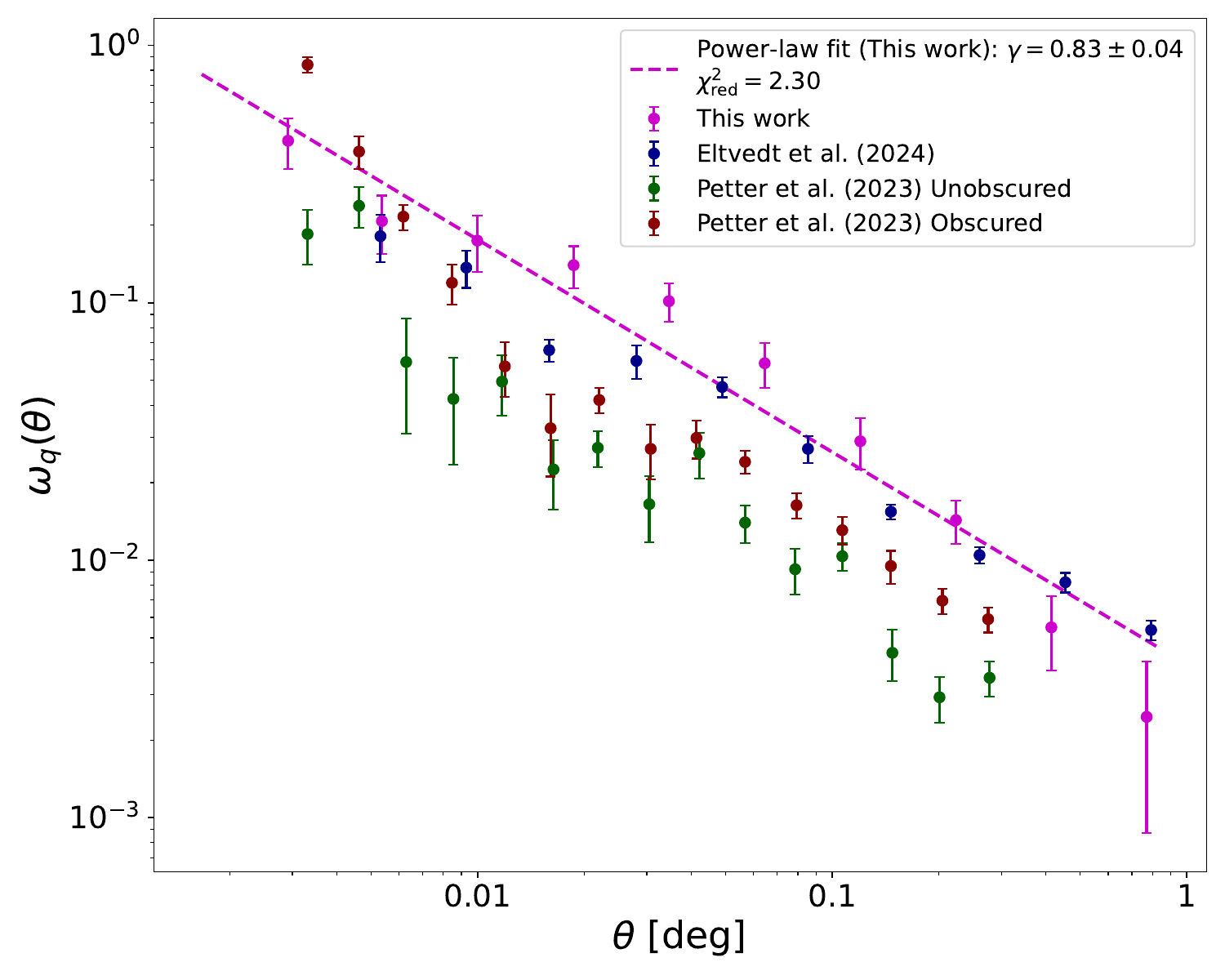}
    \caption{{Comparison of the quasar angular auto correlation function, $\omega_{q}(\theta)$, measured in this work (magenta circles) with the \cite{Eltvedt2024} measurements (blue circles), and the obscured (dark red circles) and unobscured (green circles) quasar samples from \cite{Petter2023}.  The solid magenta line represents the best-fit power-law model, $\omega_q(\theta)=A\,\theta^{-\delta}$, over the range $0.002 < \theta < 1.0$ deg. The fit yields a slope $\delta = 0.83 \pm 0.04$ with a reduced $\chi^2$ of 2.30.} }
    \label{fig:petter_elvedt}
\end{figure}
\FloatBarrier
{We compare our measurements of the angular correlation function with those of \cite{Petter2023} and \cite{Eltvedt2024}, who used a similar angular correlation technique. 
Since our main analysis is performed in tomographic redshift bins and over a different angular range (see Sec.\ref{sec:clustering}), we construct an additional quasar sample spanning the full redshift range $0.1 \leq z \leq 2.7$ to enable a direct comparison with these works. For this purpose, we measure $\omega_{q}$ over angular scales $[0.002,\,1.0]$ deg, matching as closely as possible the setup of previous studies.}
{The comparison is shown in Fig.~\ref{fig:petter_elvedt}. 
The differences in amplitude may arise from differences in quasar sample selection, redshift distributions, and the treatment of systematic effects.}

{In addition, we fit a power-law model,
\begin{equation}
\omega_q(\theta) = A\,\theta^{-\gamma}
\end{equation}
Over the angular range $0.002 < \theta < 1.0$ deg, we obtain a best-fitting slope of $\gamma = 0.83 \pm 0.04$ with a reduced chi-square of $\chi^2_{\rm red}=2.3$. The measured slope is fully consistent with the canonical value of $\gamma \simeq 0.8$ commonly found for galaxy angular correlation functions \citep{Peebles1980}.
The large reduced chi-square value indicates that a single power law is not a fully adequate description of our angular clustering measurements. }
\end{appendix}

\end{document}